\documentclass[zpreprint,zbstdefault]{./LaTeX/zeus/zeus_paper}

\usepackage[english]{babel}

\newcommand{\DO}{{D{\O}}\xspace}

\chardef\usc=95
\chardef\til=126
\catcode`\@=11 
\DeclareRobustCommand\xdotspace{\futurelet\@let@token\@xdotspace}
\def\@xdotspace{%
  \ifx\@let@token.\else
  \ifx\@let@token\bgroup.\else
  \ifx\@let@token\egroup.\else
  \ifx\@let@token\/.\else
  \ifx\@let@token\ .\else
  \ifx\@let@token~.\else
  \ifx\@let@token!.\else
  \ifx\@let@token,.\else
  \ifx\@let@token:.\else
  \ifx\@let@token;.\else
  \ifx\@let@token?.\else
  \ifx\@let@token/.\else
  \ifx\@let@token'.\else
  \ifx\@let@token).\else
  \ifx\@let@token-.\else
  \ifx\@let@token\@xobeysp.\else
  \ifx\@let@token\space.\else
  \ifx\@let@token\@sptoken.\else
   .\space
   \fi\fi\fi\fi\fi\fi\fi\fi\fi\fi\fi\fi\fi\fi\fi\fi\fi\fi}
\catcode`\@=12 

\newcommand{\stru}[2]{%
   \relax\ifmmode\hbox{\vrule height#1 depth#2 width0pt}%
   \else\vrule height#1 depth#2 width0pt\fi}

\newcommand{\Ronum}[1]{\uppercase\expandafter{\romannumeral#1}}
\newcommand{\ronum}[1]{\expandafter{\romannumeral#1}}
\DeclareRobustCommand{\LaTeXZ}{%
  \LaTeX\kern-.05em4\kern-.1em
  {\raisebox{-0.2ex}{$\scriptstyle\text{ZEUS}$}}\xspace}

\DeclareMathAlphabet{\mathbf}{OT1}{cmr}{bx}{sl}
\newcommand{\eVdist}{\kern-0.06667em}

\newcommand{\Gev}{{\text{Ge}\eVdist\text{V\/}}}

\newcommand{\gev}{{\,\text{Ge}\eVdist\text{V\/}}}

\newcommand{\pbi}{\,\text{pb}^{-1}}

\newcommand{\Tesla}{\,\text{T}}

\newcommand{\slashfrac}[2]{%
  \raisebox{0.5ex}{\ensuremath #1}\kern-0.12em/\kern-0.08em
  \raisebox{-.8ex}{\ensuremath #2}}

\newcommand{\sqr}[3]{%
    {\vcenter{\hrule height.#3ex\hbox{\vrule width.#2ex height#1ex
     \kern#1ex\vrule width.#3ex}\hrule height.#2ex}}}

\catcode`\@=11 
\newcommand{\parenbar}{\mathpalette\p@renb@r}
\def\p@renb@r#1#2{\vbox{%
  \ifx#1\scriptscriptstyle \dimen@.7em\dimen@ii.2em\else
  \ifx#1\scriptstyle \dimen@.8em\dimen@ii.25em\else
  \dimen@1em\dimen@ii.4em\fi\fi \offinterlineskip
  \ialign{\hfill##\hfill\cr
    \vbox{\hrule width\dimen@ii}\cr
    \noalign{\vskip-.3ex}%
    \hbox to\dimen@{$\mathchar300\hfil\mathchar301$}\cr
    \noalign{\vskip-.3ex}%
    $#1#2$\cr}}}
\catcode`\@=12 

\newcommand{\CC}{{\rm CC}}

\newcommand{\IP}{{\rm I$\kern-0.01667em$P}\xspace}

\newcommand{\NC}{{\rm NC}}

\mathchardef\qsm=63
\mathchardef\pls=43
\mathchardef\mns=512
\mathchardef\plm=518
\mathchardef\eql=61
\mathchardef\smallleft=300
\mathchardef\smallright=301
\mathchardef\les=316
\mathchardef\gre=318
\mathchardef\leq=532
\mathchardef\grq=533

\catcode`\@=11 
\newcounter{pict@width}
\newcounter{pict@height}
\newlength{\pict@scale}
\newcommand{\psfigadd}[4]{%
\setcounter{pict@width}{1*\ratio{#2+\pict@scale/2}{\pict@scale}}
\setcounter{pict@height}{1*\ratio{#3+\pict@scale/2}{\pict@scale}}
\setlength{\unitlength}{\pict@scale}
\hbox to #2{\hspace{-\fill}\begin{picture}(\thepict@width,\thepict@height)
\put(0,0){\psfig{figure=#1,width=#2,height=#3,clip=}}
\SetScale{0.283466457}
\SetWidth{1.763889}
{#4}
\end{picture}}
}
\newcounter{pict@widthfst}
\newcounter{pict@widthscd}
\newcounter{pict@widthtot}
\newcommand{\psfigaddtwo}[7]{%
\setcounter{pict@widthfst}{1*\ratio{#2+\pict@scale/2}{\pict@scale}}
\setcounter{pict@widthscd}{1*\ratio{#2+#4+\pict@scale/2}{\pict@scale}}
\setcounter{pict@widthtot}{1*\ratio{#2+#4+#6+\pict@scale/2}{\pict@scale}}
\setcounter{pict@height}{1*\ratio{#3+\pict@scale/2}{\pict@scale}}
\setlength{\unitlength}{\pict@scale}
\hbox{\hspace{-\fill}\begin{picture}(\thepict@widthtot,\thepict@height)
\put(0,0){\psfig{figure=#1,width=#2,height=#3,clip=}}
\put(\thepict@widthscd,0){\psfig{figure=#5,width=#6,height=#3,clip=}}
\SetScale{0.283466457}
\SetWidth{1.763889}
{#7}
\end{picture}}
}
\newcommand{\psfigror}[4]{%
\setcounter{pict@width}{1*\ratio{#2+\pict@scale/2}{\pict@scale}}
\setcounter{pict@height}{1*\ratio{#3+\pict@scale/2}{\pict@scale}}
\setlength{\unitlength}{\pict@scale}
\hbox{\begin{picture}(\thepict@width,\thepict@height)
\put(0,\thepict@height){\psfig{figure=#1,width=#3,height=#2,clip=,angle=270}}
\SetScale{0.283466457}
\SetWidth{1.763889}
{#4}
\end{picture}}
}
\newcommand{\psfigrol}[4]{%
\setcounter{pict@width}{1*\ratio{#2+\pict@scale/2}{\pict@scale}}
\setcounter{pict@height}{1*\ratio{#3+\pict@scale/2}{\pict@scale}}
\setlength{\unitlength}{\pict@scale}
\hbox{\begin{picture}(\thepict@width,\thepict@height)
\put(0,0){\psfig{figure=#1,width=#3,height=#2,clip=,angle=90}}
\SetScale{0.283466457}
\SetWidth{1.763889}
{#4}
\end{picture}}
}
\catcode`\@=12 
\newlength\listtextwidth

\catcode`\@=11 
\newlength{\@tabfninsert}
\newlength{\@tabfnwidth}
\newcommand{\tabfootnote}[2]{%
  \setlength{\@tabfninsert}{0.8em}
  \setlength{\@tabfnwidth}{\textwidth}
  \addtolength{\@tabfnwidth}{-\@tabfninsert}
  \addtolength{\@tabfnwidth}{-0.4em}
  \noindent\makebox[\@tabfninsert][r]{\footnotesize$^{#1}$\hfil}\hfill%
  \parbox[t]{\@tabfnwidth}{\footnotesize #2\hfill}}
\catcode`\@=12 

\newcommand{\ptmiss}{{\mbox{$\not\hspace{-.55ex}{P}_T$}}}
\newcommand{\rpviolate}{{\mbox{$\not\hspace{-.55ex}{R}_p$}}}

\def\citeCTD{{\cite{%
nim:a279:290,*npps:b32:181,*nim:a338:254%
}}\xspace}
\def\citeCAL{{\cite{%
nim:a309:77,*nim:a309:101,*nim:a321:356,*nim:a336:23%
}}\xspace}
\def\citeLumi{{\cite{%
desy-92-066,*zfp:c63:391,*acpp:b32:2025%
}}\xspace}
\def\cited0lq{{\cite{%
DO_1stlq_eejets,*DO_1stlq_results,*DO_1stlq_sum,*DO_1stlq_nunujets%
}}\xspace}
\def\citezeusccpaper{{\cite{%
epj:c12:411,*pl:b539:197%
}}\xspace}

\begin{document}
\prepnum{{DESY--03--041}}

\title{
A search for resonance decays to lepton+jet at HERA \\
and limits on leptoquarks
}                                                       
                    
\author{ZEUS Collaboration}
\draftversion{3.6}
\date{March 2003}

\abstract{
A search for narrow-width resonances that decay into electron+jet 
or neutrino+jet has been performed with the ZEUS
detector at HERA operating at
center-of-mass energies of 300 and 318~$\gev$.
An integrated $e^+p$ luminosity of
114.8~$\pbi$ and $e^-p$ luminosity of 
16.7~$\pbi$ were used.
No evidence for any resonance was found. Limits were derived
on the Yukawa coupling, $\lambda$, as a function of
the mass of a hypothetical resonance that 
has arbitrary decay branching ratios into $eq$ or $\nu q$.
These limits also apply to squarks predicted by
$R$-parity-violating supersymmetry.
Limits for the production of leptoquarks
described by the
Buchm$\ddot{\rm{u}}$ller-R$\ddot{\rm{u}}$ckl-Wyler model
were also derived for masses up to 400~$\gev$. 
For $\lambda$~=~0.1, leptoquark masses
up to 290~GeV are excluded.
}

\makezeustitle

\def\3{\ss}                                                                                        
\newcommand{\address}{ }                                                                           
\newcommand{\author}{ }                                                                          
\pagenumbering{Roman}                                                                              
                                                                              
                                                   %
\begin{center}                                                                                     
{                      \Large  The ZEUS Collaboration              }                               
\end{center}                                                                                       
  S.~Chekanov,                                                                                     
  M.~Derrick,                                                                                      
  D.~Krakauer,                                                                                     
  J.H.~Loizides$^{   1}$,                                                                          
  S.~Magill,                                                                                       
  B.~Musgrave,                                                                                     
  J.~Repond,                                                                                       
  R.~Yoshida\\                                                                                     
 {\it Argonne National Laboratory, Argonne, Illinois 60439-4815}~$^{n}$                            
\par \filbreak                                                                                     
  M.C.K.~Mattingly \\                                                                              
 {\it Andrews University, Berrien Springs, Michigan 49104-0380}                                    
\par \filbreak                                                                                     
  P.~Antonioli,                                                                                    
  G.~Bari,                                                                                         
  M.~Basile,                                                                                       
  L.~Bellagamba,                                                                                   
  D.~Boscherini,                                                                                   
  A.~Bruni,                                                                                        
  G.~Bruni,                                                                                        
  G.~Cara~Romeo,                                                                                   
  L.~Cifarelli,                                                                                    
  F.~Cindolo,                                                                                      
  A.~Contin,                                                                                       
  M.~Corradi,                                                                                      
  S.~De~Pasquale,                                                                                  
  P.~Giusti,                                                                                       
  G.~Iacobucci,                                                                                    
  A.~Margotti,                                                                                     
  R.~Nania,                                                                                        
  F.~Palmonari,                                                                                    
  A.~Pesci,                                                                                        
  G.~Sartorelli,                                                                                   
  A.~Zichichi  \\                                                                                  
  {\it University and INFN Bologna, Bologna, Italy}~$^{e}$
% ...
\newpage                                                                                           
$^{\    1}$ also affiliated with University College London \\                                      
% ...
                                                           %
                                                           %
\newpage   
                                                           %
                                                           %
\begin{tabular}[h]{rp{14cm}}                                                                       
$^{a}$ &  supported by the Natural Sciences and Engineering Research                               
          Council of Canada (NSERC) \\                                                             
$^{b}$ &  supported by the German Federal Ministry for Education and                               
          Research (BMBF), under contract numbers HZ1GUA 2, HZ1GUB 0, HZ1PDA 5, HZ1VFA 5\\         
$^{c}$ &  supported by the MINERVA Gesellschaft f\"ur Forschung GmbH, the                          
          Israel Science Foundation, the U.S.-Israel Binational Science                            
          Foundation and the Benozyio Center                                                       
          for High Energy Physics\\                                                                
$^{d}$ &  supported by the German-Israeli Foundation and the Israel Science                        
          Foundation\\                                                                             
$^{e}$ &  supported by the Italian National Institute for Nuclear Physics (INFN) \\                
$^{f}$ &  supported by the Japanese Ministry of Education, Science and                             
          Culture (the Monbusho) and its grants for Scientific Research\\                          
$^{g}$ &  supported by the Korean Ministry of Education and Korea Science                          
          and Engineering Foundation\\                                                             
$^{h}$ &  supported by the Netherlands Foundation for Research on Matter (FOM)\\                   
$^{i}$ &  supported by the Polish State Committee for Scientific Research,                         
          grant no. 620/E-77/SPUB-M/DESY/P-03/DZ 247/2000-2002\\                                   
$^{j}$ &  partially supported by the German Federal Ministry for Education                         
          and Research (BMBF)\\                                                                    
$^{k}$ &  supported by the Fund for Fundamental Research of Russian Ministry                       
          for Science and Edu\-cation and by the German Federal Ministry for                       
          Education and Research (BMBF)\\                                                          
$^{l}$ &  supported by the Spanish Ministry of Education and Science                               
          through funds provided by CICYT\\                                                        
$^{m}$ &  supported by the Particle Physics and Astronomy Research Council, UK\\                   
$^{n}$ &  supported by the US Department of Energy\\                                               
$^{o}$ &  supported by the US National Science Foundation\\                                        
$^{p}$ &  supported by the Polish State Committee for Scientific Research,                         
          grant no. 112/E-356/SPUB-M/DESY/P-03/DZ 301/2000-2002, 2 P03B 13922\\                    
$^{q}$ &  supported by the Polish State Committee for Scientific Research,                         
          grant no. 115/E-343/SPUB-M/DESY/P-03/DZ 121/2001-2002, 2 P03B 07022\\                    
\end{tabular}                                                                                      
                                                           %
                                                           %

\clearpage
\pagenumbering{arabic} 
\pagestyle{plain}




\section{Introduction}

Many extensions of the Standard Model (SM)
predict the existence of
particles carrying both baryon and lepton numbers, such as
leptoquarks (LQs) \cite{pl:b191:442} or squarks 
(in $R$-parity-violating
($\rpviolate$) supersymmetry)~\cite{np:b397:3}.
In $ep$ collisions at HERA, 
such states may be produced directly through
electron\footnote{Unless otherwise specified,
`electron' refers to both positron and electron
and `neutrino' refers to both neutrino and antineutrino.}-quark
fusion, 
with subsequent decay into  
electron and quark or neutrino and quark,
yielding peaks in the spectra of
the final-state lepton-jet invariant mass, $M_{lj}$.
This paper presents a search for such resonant states.

\begin{figure}[tbph]
\begin{center}
\epsfig{figure=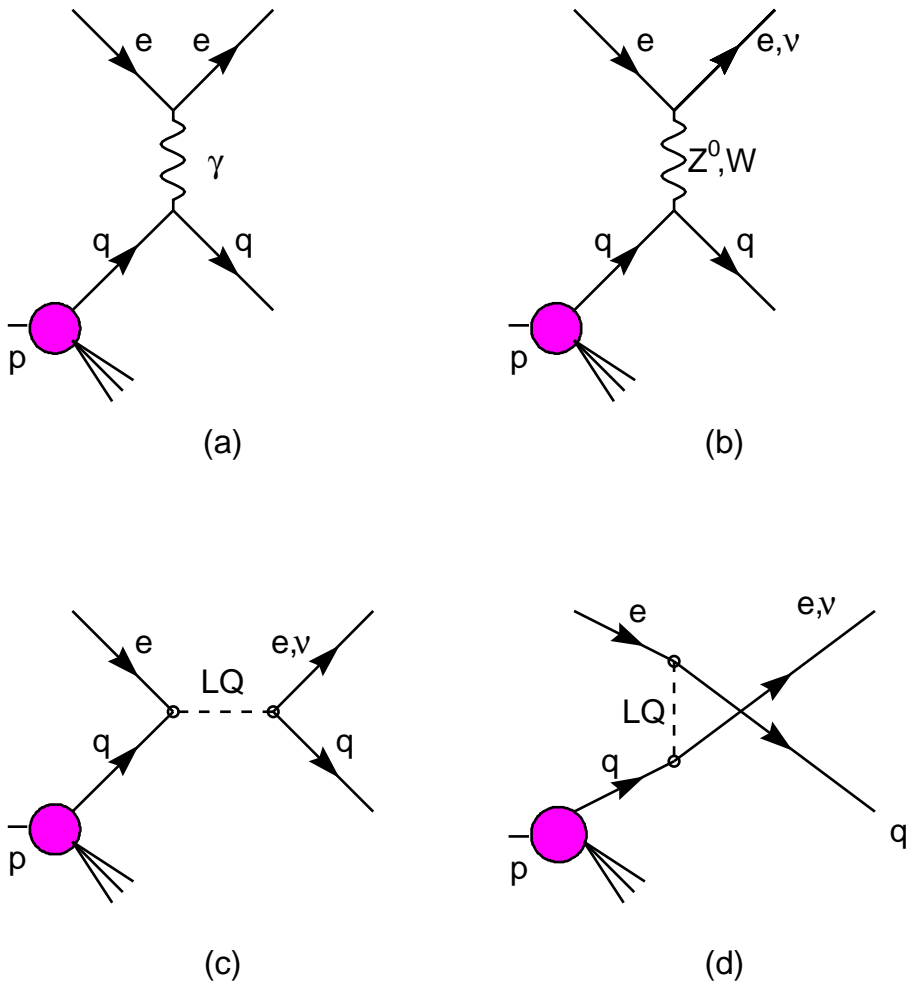,height=16.cm}
\end{center}
\caption{Diagrams for $ep$ scattering at HERA
via (a) photon and (b) $Z^0$ exchange (NC) and via
$W$ exchange (CC). The leptoquark diagrams for
the same initial and final states are c) $s$-channel
LQ production and d) $u$-channel LQ exchange.
Here $e$ stands for both $e^+$ (positron) and $e^-$
(electron), and $\nu$ for both
$\nu$ (neutrino) and $\bar\nu$ (antineutrino).}
\label{lq_feynman}
\end{figure}

The only significant backgrounds
to high-mass resonance production
arise from neutral current (NC) 
and charged current (CC) deep inelastic scattering (DIS), 
as illustrated in Figs.~\ref{lq_feynman}a and b.
While a resonance with mass below 
the HERA center-of-mass energy, 
$\sqrt{s}$,
would give rise to narrow peak in the
$M_{lj}$ spectrum, the backgrounds from NC and CC
fall rapidly at high mass due to 
the dependence of the cross section on $Q^2$,
the virtuality of the exchanged bosons,
and to the
sharply falling valence-quark density at large Bjorken $x$.
The variable $\theta^*$,
the lepton scattering 
angle in the lepton-jet center-of-mass frame,
can be used to reduce
the DIS backgrounds.  The decay of a resonance
results in an angular  
distribution different from those
produced by SM processes: 
a scalar resonance, for example, will have a
flat distribution in $\cos \theta^*$, while NC DIS events
follow approximately a $1/(1-\cos\theta^*)^2$ distribution.

The data were used to investigate the production of resonances. 
In the absence of a narrow resonance signal, limits can
be set on the production of resonances with
masses below $\sqrt{s}$ and with the assumption
of narrow width, as shown in Fig.~\ref{lq_feynman}c.
In addition, specific limits
on the production of 
Buchm$\ddot{\rm{u}}$ller-R$\ddot{\rm{u}}$ckl-Wyler (BRW) LQs
\cite{pl:b191:442}
with masses both below and above 
$\sqrt{s}$ can be obtained.  
Because backgrounds fall sharply at large $M_{lj}$, 
the data are sensitive to LQs via 
exchange terms, shown in Fig.~\ref{lq_feynman}d,
as well as interference terms with SM processes.
The narrow-width assumption is not necessary for setting 
limits in the BRW model.

The present results supersede previous analyses
\cite{epj:c16:253,pr:d63:052002}. They are based on all
data collected by the ZEUS experiment in the period from 1994
to 2000.  After 1998, the HERA center-of-mass energy was increased from
300 to 318~$\gev$.

\section{Experimental conditions}

The ZEUS detector is described in detail elsewhere
\cite{zeus:1993:bluebook}. 
The main components used in the present analysis are the central
tracking detector (CTD)
and the uranium-scintillator sampling calorimeter (CAL). 

Charged particles are tracked in the CTD~\citeCTD,
which operates in a magnetic field of $1.43\Tesla$ provided by a thin
superconducting solenoid. The CTD consists of 72~cylindrical drift chamber
layers, organized in nine superlayers covering the polar-angle region
{$15^\circ<\theta<164^\circ$}. The transverse-momentum resolution for
full-length tracks is 
$\sigma(p_T)/p_T=0.0058p_T\oplus0.0065\oplus0.0014/p_T$,
with $p_T$ in $\Gev$.

The CAL~\citeCAL consists 
of three parts: the forward\footnote{The 
ZEUS coordinate system is a right-handed Cartesian system,
with the $Z$ 
axis pointing in the proton beam direction,
referred to as the ``forward direction'',
and the $X$ axis 
pointing left towards the center 
of HERA. The coordinate origin is at the nominal interaction point.
}
(FCAL), the barrel (BCAL) 
and the rear (RCAL)
calorimeters. Each part is divided into modules, which
are subdivided transversely into towers and
longitudinally into one electromagnetic section (EMC) and 
either one (in RCAL)
or two (in BCAL and FCAL) hadronic sections (HAC). 
The smallest subdivision of
the calorimeter is called a cell.  The CAL energy resolutions, 
as measured under
test-beam conditions, are $\sigma(E)/E=0.18/\sqrt{E}$ 
for electrons and
$\sigma(E)/E=0.35/\sqrt{E}$ for hadrons, with $E$ in $\Gev$.
The timing resolution of the CAL is better than 1 ns for energy
deposits greater than 4.5 $\gev$.
Further performance parameters of the CAL
relevant for this study have been discussed
in previous publications~\cite{epj:c16:253,pr:d63:052002}.

The forward
plug calorimeter (FPC), a lead-scintillator
sandwich calorimeter, was installed in the 20 $\times$ 20 cm$^2$
beam hole of the forward CAL (FCAL) of the ZEUS detector in 1998.
Although the FPC information was not used in this analysis,
the impact of its material on the CAL response to forward jets was
extensively studied.

The luminosity, which was measured~\citeLumi
from the rate of the bremsstrahlung
process $ep \rightarrow ep \gamma$,
has an uncertainty of 1.6\% to 2.25\%, depending
on the running periods.
All ZEUS data
collected from 1994 to 2000 are listed in
Table \ref{tab:lumi}, and were used for this analysis.

\section{Monte Carlo simulation}
\label{mc_simulation}

Production and decay of resonances 
were simulated using PYTHIA 6.1~\cite{cpc:135:238,pythia_lq61_hera},
which takes into account the finite width of
the resonant state, but includes only the $s$-channel diagrams. 
Initial-
and final-state QCD radiation from the quark and the effect of LQ 
hadronization~\cite{pythia_lq61_hera} 
before decay as well as the initial-state
QED radiation from the electron are also taken into account.

Standard Model NC and CC DIS events were simulated using the 
HERACLES 4.6.2~\cite{cpc:69:155} program 
with the DJANGO 6 version 2.4~\cite{cpc:81:381}
and DJANGOH 1.1~\cite{spi:www:djangoh11}
interfaces to the hadronization programs.
Radiative corrections for initial- and final-state 
electroweak radiations, vertex and propagator corrections,
and two-boson exchange were included.
The hadronic final state was simulated using the MEPS model in LEPTO 
6.5~\cite{cpc:101:108}, which includes $O$($\alpha_S$) matrix elements
and higher-order QCD radiation.
The CTEQ5D parton distribution function (PDF)~\cite{pr:d55:1280} 
was used in evaluating the SM cross sections.

The largest uncertainty in the 
NC and CC cross sections is due to the uncertainties in
the parton densities of the proton. The PDFs
at high Bjorken $x$ are determined primarily 
from measurements made in fixed-target DIS 
experiments.  At $x=0.6$, corresponding to a 
lepton-jet mass of $230\gev$, 
the cross-section uncertainty due to the 
PDF uncertainty~\cite{epj:c14:285}
is $\approx6\%$ for NC and $\approx4$ (10)\%
for $e^-u$ ($e^+d$) CC reactions,
where $u$ and $d$ refer to the $u$ and $d$ 
valence quarks, respectively.

The generated events were passed through
the GEANT 3.13-based \cite{tech:cern-dd-ee-84-1}
ZEUS detector- and trigger-simulation programs
~\cite{zeus:1993:bluebook}.
They were reconstructed and analysed by the same program
chain as the data.

\section{Resonance search}

Events from a hypothetical
resonance decaying
into $eq$ ($\nu$$q$) have a
topology identical to DIS $\NC$ ($\CC$) events.
Events originating from 
high-mass resonances are expected to have 
high transverse energy, at
least one jet, and either an identified final-state electron
or large missing transverse energy.
The lepton-jet invariant mass was calculated as
\begin{equation}
\notag
M_{lj} = \sqrt{2E_lE_j(1-\cos\xi)} ,
\end{equation}
where $E_l$ is the energy of the outgoing lepton, 
$E_j$ is
the energy of the jet, and $\xi$ is the angle
between the lepton and jet.
In final states containing multiple jets,
the jet with the largest transverse momentum, $P_{T}^j$, was used.

\subsection{$ep \to eX$ topology}
\label{nc_measurement}

\subsubsection{Event selection}
\label{sec:event_select_nc}

Events with the topology $ep \to eX$, where
$X$ contains one or more jets,
were selected using the following criteria:

\begin{itemize}
\item the $Z$ coordinate
of the reconstructed event 
vertex was required to be in the
range $|Z|<50$~cm, consistent with an $ep$ collision;

\item  the total transverse energy, $E_{T}$, was
required to be at least $60\gev$. This removes the bulk of the SM NC
background;

\item  an identified electron
\cite{zfp:c74:207} was required
with energy $E_{e}^{\prime }>25$~GeV located
either in FCAL or BCAL, 
corresponding to an electron polar angle,
$\theta_e < 126^{\circ}$. 
Electrons impacting the BCAL within $1.5$~cm of a module
edge, as well as electrons impacting between the FCAL and the BCAL,
as determined by 
tracking information, were discarded to ensure
that the resolutions were well understood;

\item  at least one hadronic jet with transverse momentum 
$P^{j}_{T}>15\gev$, obtained
using the longitudinally invariant
$k_T$ cluster algorithm~\cite{np:b406:187} in
inclusive mode~\cite{pr:d48:3160}, was required.
The centroid at the FCAL face of the highest-$P_T$ jet
was required to be outside a box of
$60\times 60$~cm$^2$ centered on the proton beam,
in order to ensure good energy containment and to reduce
the systematic uncertainties due to the proton remnant.

\end{itemize}

The acceptance,
mass shifts, and resolutions for resonant lepton-quark states 
were calculated from the LQ MC.
After these cuts, the acceptance for scalar resonances was 
$\sim$60\%, depending weakly on the mass.
The acceptance for vector resonances was $\sim$60\% below
200~$\gev$, decreasing to $\sim$40\% at 290~$\gev$.
These differences in the acceptances for scalar and
vector resonances are
due to the different decay angular distributions.
The mass resolution,
determined from a Gaussian fit to the peak of the reconstructed mass
spectrum, fell from 6\% to 4\% as the resonant mass increased
from 150 to $290\gev$. The peak position 
was typically lower than the generated mass by 1\%.
The resolution
in $\cos \theta ^{\ast }$ near $|\cos \theta ^{\ast }|=1$ had 
a Gaussian width 
of $0.01$, degrading to $0.03$ with decreasing
$|\cos \theta ^{\ast }|$.

\subsubsection{Search results}

\begin{figure}[tbph]
\begin{center}
\epsfig{figure=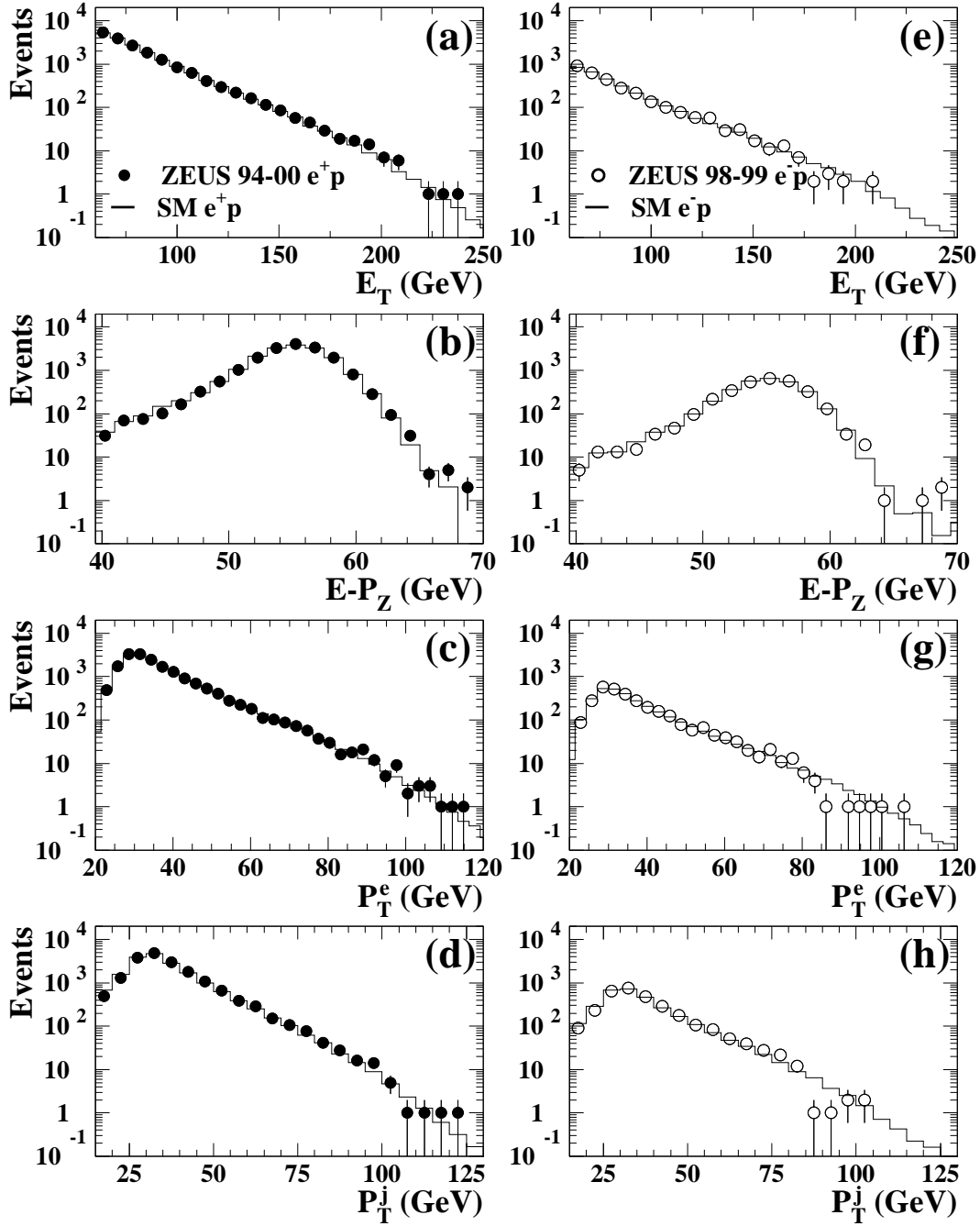,height=20.cm}
\end{center}
\caption{Comparison of distributions from 94-00 $e^+p$ (a-d) 
and from 98-99 $e^-p$ data sets (e-h)
with the corresponding NC MC samples: 
(a,e) total transverse energy, $E_{T}$; 
(b,f) $E-P_Z$; (c,g) electron tranverse momentum, $P_{T}^{e}$, and
(d,h) jet transverse momentum, $P_{T}^{j}$.}
\label{nc_control}
\end{figure}

After the above selection, 21\,509 events were found, 
compared to 21\,445 $\pm$ 1\,288
expected from the $\NC$ MC simulation
and the evaluation of its systematic uncertainties
(see Section~\ref{sec:nc_systematic}).
The measured distributions of the total
transverse energy, $E_{T}$, are compared to the
simulation in Figs.~\ref{nc_control}a and \ref{nc_control}e,
where the $e^+p$ and $e^-p$ samples are shown separately.
Also shown
are $E-P_Z$ (Figs.~\ref{nc_control}b,f), 
where the $E$ and $P_Z$ are summed over the
final-state electron and all the hadrons,
electron transverse momentum
($P_{T}^{e}$) (Figs.~\ref{nc_control}c,g) and
jet transverse momentum ($P_{T}^{j}$)
(Figs.~\ref{nc_control}d,h).
Good agreement is seen between the
data and the SM NC simulation for all of these spectra.

\begin{figure}[tbph]
\begin{center}
\epsfig{figure=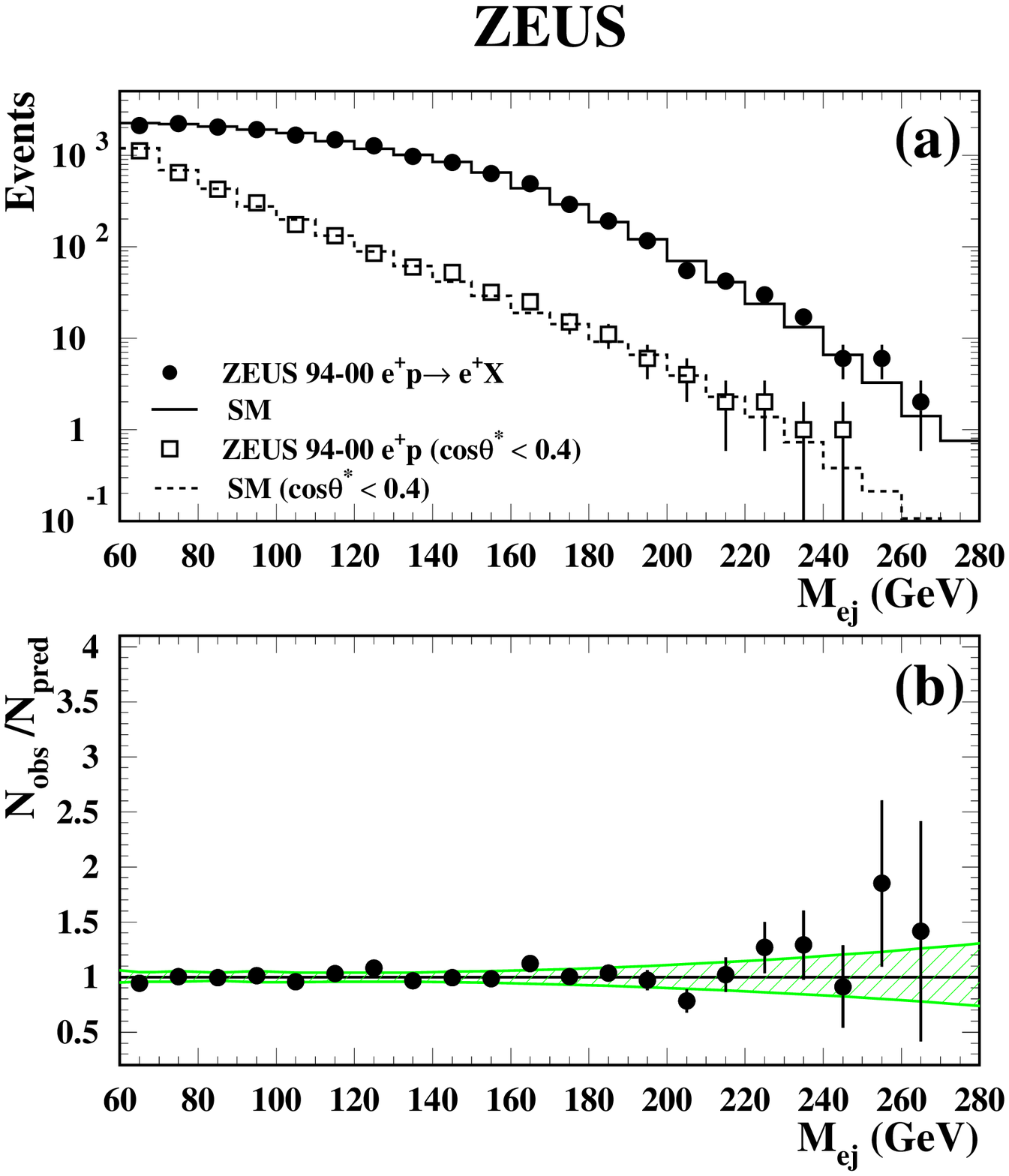,height=16.cm}
\end{center}
\caption{(a) Comparison of 
the $e^+p$ samples (dots) and the NC SM 
expectations (solid histogram) for the 
reconstructed invariant mass $M_{ej}$ in
the $e^+p \to e^+X$ topology.
The data (open squares) and the SM expectations (dashed
histogram) after the $\cos\theta^*<$0.4 cut are also shown.
(b) The ratio between the data and the SM expectation before the
$\cos\theta^*$ cut.
The shaded area shows the overall uncertainties of
the SM MC expectation.}
\label{nc_mass_posi}
\end{figure}

\begin{figure}[tbph]
\begin{center}
\epsfig{figure=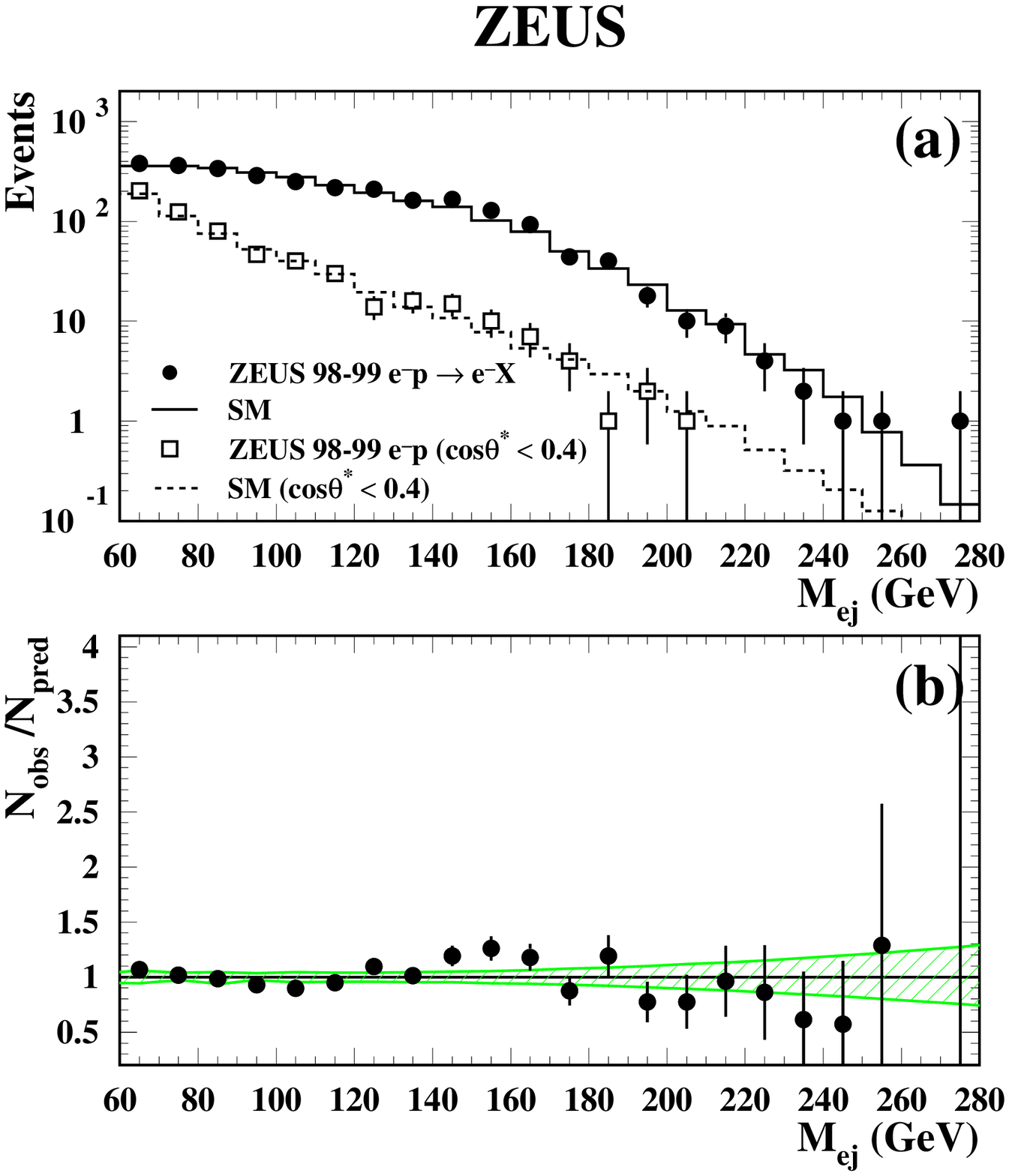,height=16.cm}
\end{center}
\caption{(a) Comparison of
the observed $e^-p$ samples (dots) and the NC SM 
expectations (solid histogram) for the 
reconstructed invariant mass $M_{ej}$ in
the $e^-p \to e^-X$ topology.
The data (open squares) and the SM expectations (dashed
histogram) after the $\cos\theta^*<$0.4 cut are also shown.
(b) The ratio between the data and the SM expectation before the
$\cos\theta^*$ cut.
The shaded area shows the overall uncertainties on
the SM MC expectation.}
\label{nc_mass_elec}
\end{figure}

Figures \ref{nc_mass_posi} and \ref{nc_mass_elec}
show the 
$M_{ej}$ spectra for $e^+p$ and $e^-p$ data, respectively.
The upper plots
show the spectra with and without the cut
$\cos\theta^*<0.4$, 
while the lower plots show the ratio
of the observed spectrum to SM expectations with no $\cos\theta^*$
cut.  The data are well described by the NC MC.

An excess of data events relative to SM expectations
was seen in data from the 1994-97 period~\cite{epj:c16:253}.
For $M_{ej}>210\gev$,
24.7 events were expected and 49 were observed.
No such excess is seen in the more recent data.
The increase in the proton beam energy from 820 
to 920 $\gev$ in 1998 had no
significant effect on the mass reconstruction
or the signal acceptance.
The effect of the addition of the FPC was 
studied extensively;
uncertainties in its simulation were found
not to affect the conclusions of the present analysis.
Combining all the $e^+p$ data, 
104 events were observed  with $M_{ej}>210\gev$,
in good agreement
with the SM expectation of 90 $\pm 15$.  
The systematic uncertainty on the expectation
is discussed in the next section. 
The excess seen in the earlier data sample 
must be ascribed to
a statistical fluctuation.

\subsubsection{Systematic uncertainties}
\label{sec:nc_systematic}

The uncertainty
on the expected number of events from NC DIS process
was investigated.
The dominant sources were uncertainties in:

\begin{itemize}
\item calorimeter energy scale, of
1\% for BCAL electrons, 2\% for FCAL electrons 
and 2\% for hadrons.
This led to an uncertainty of
$4~(12)$\% in the NC expectation for $M_{ej}$=100 (220)$\gev$;

\item the simulation of the hadronic energy flow
including simulation of the proton remnant and the energy flow
between the struck quark and proton remnant.
Tests based on the SM MC samples
yielded 
variations of the NC background of about
10\% for masses above 220 $\gev$;

\item the energy reponse
of FCAL towers closest to the beam
from the differences observed between data and simulation.
Tests based on the SM MC samples showed variations
of the NC background of less than 5\% for $M_{ej}$=220 $\gev$;

\item the parton densities, as
estimated by Botje \cite{epj:c14:285}, which gave an uncertainty of 5\%
for $M_{ej}$=220 $\gev$.


\end{itemize}

Other uncertainties were investigated
and found to be small compared to the above items.
They are the simulation of the
electron-energy resolution,
the electron-finding efficiency,
the jet-position reconstruction,
the luminosity determination and the simulation
of the vertex distribution.
The overall systematic uncertainties 
on the background expectations result from
summing the contributions from all these sources in quadrature and are
shown in Figs.~\ref{nc_mass_posi}b and \ref{nc_mass_elec}b
as the shaded bands.

\subsubsection{Significance analysis}

To quantify the level of agreement of the
$M_{ej}$ spectra between the SM MC and the data,
a significance analysis was performed
using a sliding mass window of
width 3$\sigma(M_{ej})$, where $\sigma(M_{ej})$ is the mass
resolution discussed in Section \ref{sec:event_select_nc}.  
The number of events, $\mu$, expected from
the SM and the number of observed events, $N$,
were compared in each window, and the
probability of observing $N$ or more events
while expecting $\mu$ was calculated as 
\begin{equation}
P=\sum_{k=N}^{\infty}e^{-\mu}
              \frac{{\mu}^k}{k!}.
\label{poisson_more}
\end{equation}
In the $e^+p$ data sample,
a minimum probability, $P_{min}$,
of 2.6$\times$10$^{-3}$ was found at mass 121~$\gev$, where
2\,575 events were observed while 2\,436 were expected
inside the sliding window.
A large number of MC experiments were then performed, taking into
account the systematic uncertainties on the SM expectations,
and the $P_{min}$ distribution was determined.
In 35\% of all simulated experiments, the value of
$P_{min}$ obtained was less than that found in the data. 
With the additional cut $\cos\theta^*<0.4$
applied to the sample, 
a $P_{min}$ of
1.9$\times$10$^{-2}$ was found in the $e^+p$ data sample at a
mass of 158~$\gev$.  A value of $P_{min}$
less than that found in the data was observed in 
41\% of the generated experiments.
The same test was also done on the $e^-p$ data samples.
The results are summarized in Table~\ref{tab:many_experiment_nc}.
These observations show that the data are 
compatible with the SM expectation, and there is no evidence for
a narrow resonance in the $eq$ channel.

\subsection{$ep \to \nu X$ topology}
\label{cc_measurement}

\subsubsection{Event selection}
Events with the topology $ep \to \nu X$,
where $X$ contains one or more jets, 
were selected by the following cuts,
similar to those used in the CC cross-section measurement
\citezeusccpaper:

\begin{itemize}

\item the $Z$ coordinate of the reconstructed event 
vertex was required to be in the range $|Z|<50$~cm.
The event vertex was reconstructed either using
the tracks measured in the CTD (for events with large
$\gamma_0$~\citezeusccpaper, 
the hadronic scattering angle of the system
relative to the nominal interaction point)
or from the arrival time of the particles entering the FCAL
(for events with small $\gamma_0$, i.e. outside
the CTD acceptance);

\item the missing transverse momentum,
$\ptmiss >20\gev$, as measured in the calorimeter;

\item $y<0.9$, where $y$
was calculated from the longitudinal momentum measured 
in the calorimeter: $y=(E-P_Z)/2E_e$, where
$E_e$=27.5$\gev$ is the electron beam energy.
This cut discards events in which the 
kinematic variables were poorly reconstructed;

\item NC events were removed by discarding events with identified
electrons;

\item at least one jet was required with 
$P_{T}^j>10\gev$, where jets were reconstructed 
as in the $ep \to eX$ topology, and 
the centroid at the FCAL face of the highest-$P_T$ jet
was required to be outside a box of
$60\times 60$~cm$^2$ centered on the proton beam.

\end{itemize}

The neutrino energy and angle were calculated 
by assuming that $\ptmiss$ and missing $E-P_Z$
were carried away by a single neutrino.
Monte Carlo simulations of resonant-state production 
indicated that the
neutrino energy, $E_{\nu}$, and polar angle, $\theta_{\nu}$,
were measured with average
resolutions of $16\%$ and $11\%$, respectively. The average systematic 
shift in $E_{\nu}$ was less than $2\%$,  while the shift in 
$\theta_{\nu}$ was less than $1\%$.
After all these cuts, the acceptance was $\sim$ 60\%,
depending weakly on the mass.

The invariant mass of the $\nu$-jet system,~$M_{\nu j}$,
was reconstructed using the sum of the
4-momenta of the 
neutrino and the highest-$P_T$ jet in the event.
The shift and resolution of the invariant mass
were studied by reconstructing it
for the LQ MC events and
fitting the mass peak with a Gaussian function.
The resulting mass shift was below
1\% for LQ masses between
150 and $290\gev$, with the resolution varying from 8\% to 7\%.

\subsubsection{Search results}

\begin{figure}[tbph]
\begin{center}
\epsfig{figure=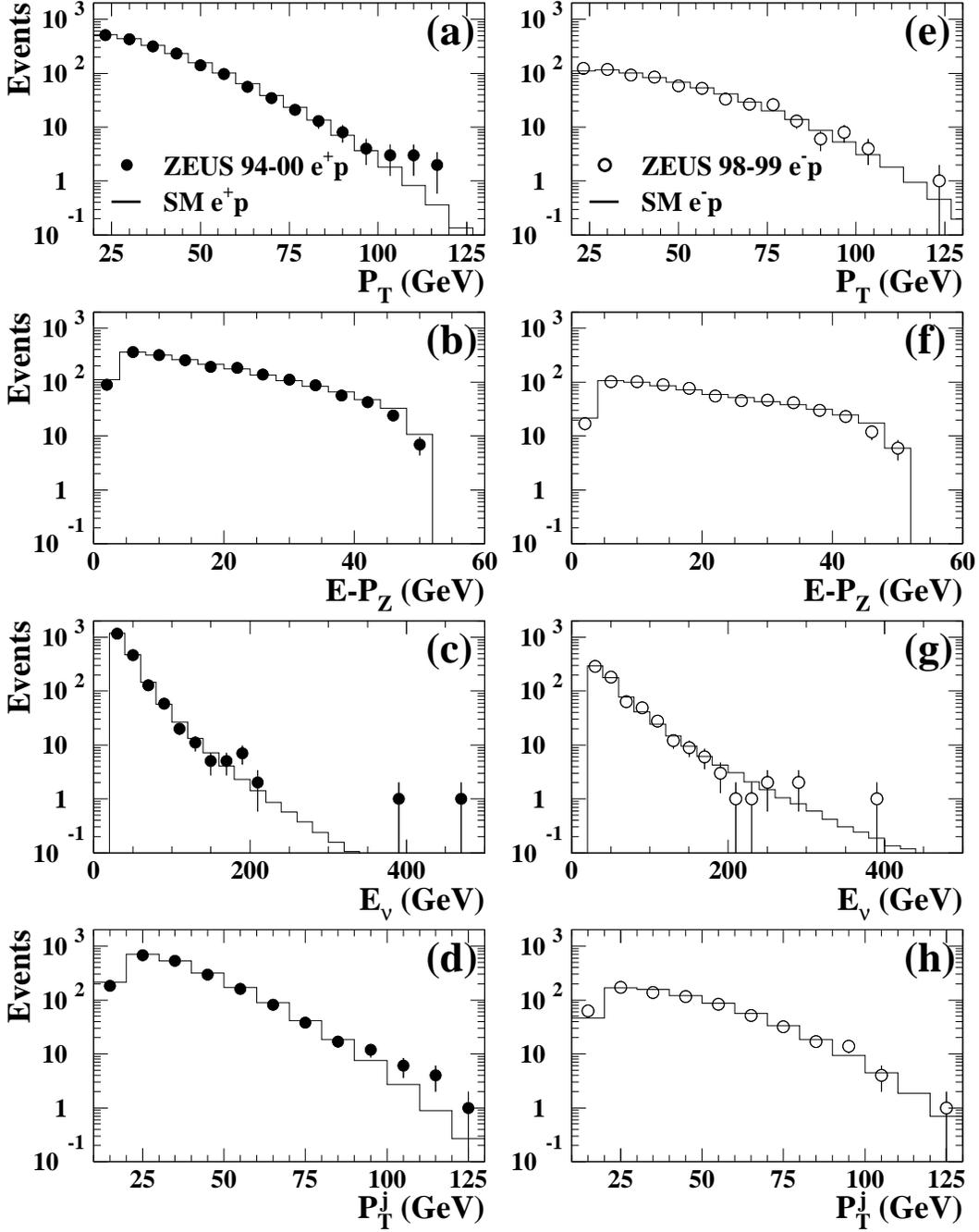,height=20.cm}
\end{center}
\caption{Comparison of distributions from 94-00 $e^+p$ (a-d)
and from 98-99 $e^-p$ data sets (e-h)
with the corresponding CC MC
samples for the selected distributions: 
(a,e) missing transverse energy, $P_{T}$; 
(b,f) $E-P_Z$; (c,g) neutrino energy, $E_{\nu}$, and
(d,h) jet transverse momentum, $P_{T}^{j}$.}
\label{cc_control}
\end{figure}

After the selection, 2\,536 events were found, compared to 2\,587
$\pm$ 217
expected from the $\CC$ MC simulation
and the evaluation of its systematic uncertainties
(see Section~\ref{sec:cc_systematic}). 
The distributions of $\ptmiss$
are compared for data and
simulation in Figs.~\ref{cc_control}a and \ref{cc_control}e,
where the $e^+p$ and $e^-p$ samples are shown separately.
Also shown
are $E-P_Z$ (Figs.~\ref{cc_control}b,f), 
where the $E$ and $P_Z$ are summed over the
final-state hadrons,
$E_{\nu}$ (Figs.~\ref{cc_control}c,g) and
$P_{T}^{j}$ (Figs.~\ref{cc_control}d,h).
Reasonable agreement is seen between the
data and the SM CC simulation for all of these spectra.
A small excess compared to MC is observed
for the $e^+p$ data at large $\ptmiss$
(also reflected in the $P_T^j$ and $E_{\nu}$ distributions).

\begin{figure}[tbph]
\begin{center}
\epsfig{figure=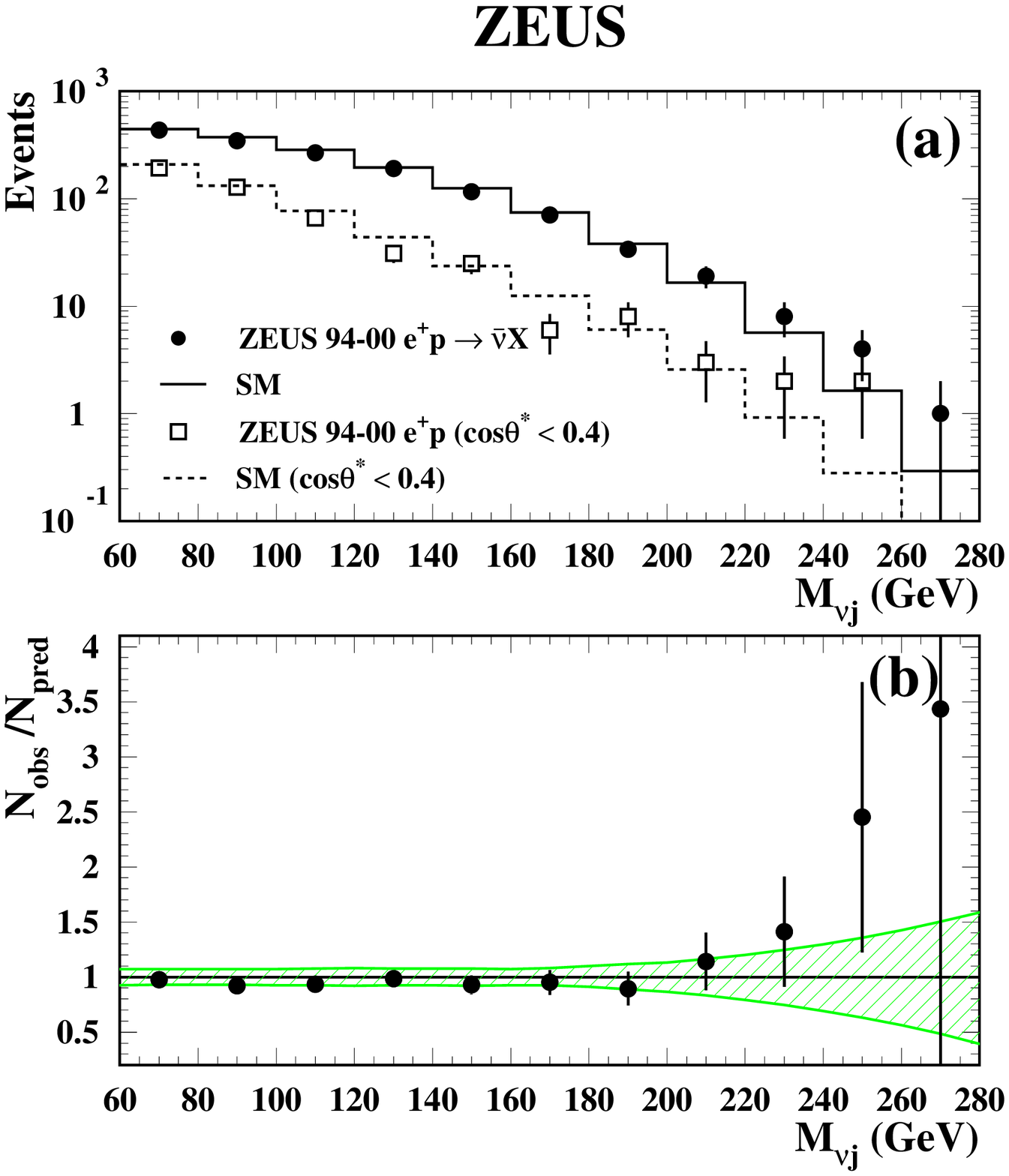,height=16.cm}
\end{center}
\caption{(a) Comparison of 
the observed $e^+p$ samples (dots) and the CC SM 
expectations (solid histogram) for the 
reconstructed invariant mass $M_{\nu j}$ in
the $e^+p \to \bar{\nu}X$ topology.
The data (open squares) and the SM expectations (dashed
histogram) after the $\cos\theta^*$$<$0.4 cut are also shown.
(b) The ratio between the data and the SM expectation before the
$\cos\theta^*$ cut.
The shaded area shows the overall uncertainties on
the SM MC expectation.}
\label{cc_mass_posi}
\end{figure}

\begin{figure}[tbph]
\begin{center}
\epsfig{figure=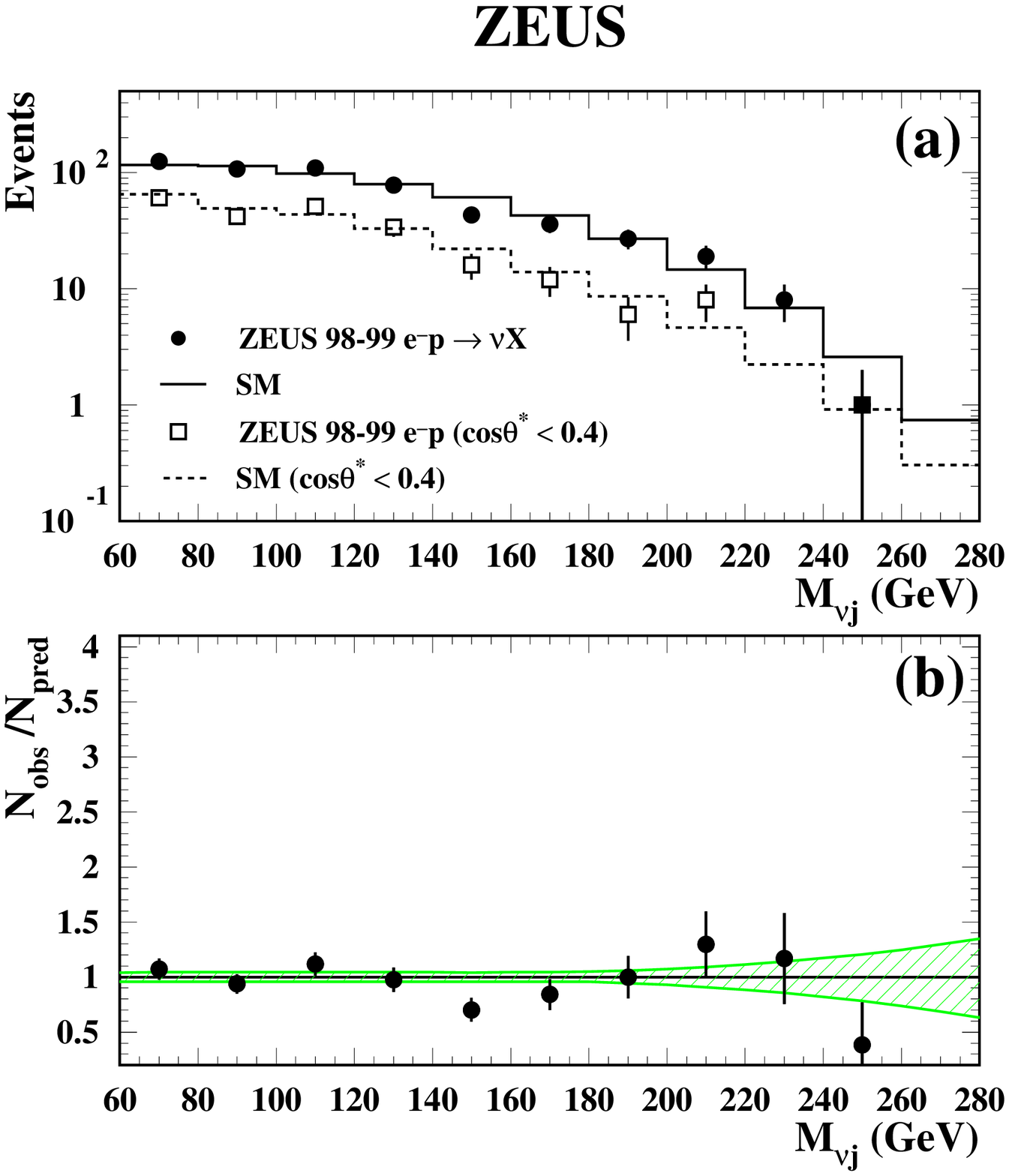,height=16.cm}
\end{center}
\caption{(a) Comparison of 
the observed $e^-p$ samples (dots) and the CC SM 
expectations (solid histogram) for the 
reconstructed invariant mass $M_{\nu j}$ in
the $e^-p \to \nu X$ topology.
The data (open squares) and the SM expectations (dashed
histogram) after the $\cos\theta^*$$<$0.4 cut are also shown.
(b) The ratio between the data and the SM expectation before the
$\cos\theta^*$ cut.
The shaded area shows the overall uncertainties on
the SM MC expectation.}
\label{cc_mass_elec}
\end{figure}

Figures \ref{cc_mass_posi} and \ref{cc_mass_elec}
show the 
$M_{\nu j}$ spectra for $e^+p$ and $e^-p$ data, respectively.
The upper plots show the spectra with and without a cut of
$\cos\theta^*<0.4$, while the lower plots show the ratio
of the observed spectra to SM expectations with no $\cos\theta^*$ 
cut. The data are reasonably well described by the CC MC.

\subsubsection{Systematic uncertainties}
\label{sec:cc_systematic}

The uncertainty on the predicted background from SM CC DIS processes
was investigated. The dominant sources of uncertainty, which
are similar to those described in Section \ref{sec:nc_systematic}
for the $ep \to eX$ case,
arise from uncertainties in:
\begin{itemize}
\item the hadronic energy scale, of 2\%, which leads to an uncertainty
of 2 (10)\% for $M_{\nu j}$=100 (220) $\gev$;

\item the simulation of 
the energy deposited in the FCAL regions
closest to the forward beam pipe, which
leads to an uncertainty of $\sim$ 7\% for $M_{\nu j}$=220 $\gev$;

\item the parton densities, as estimated by Botje
\cite{epj:c14:285}, giving 9\% and 4\% uncertainties for
$e^+p$ and $e^-p$, respectively,
for $M_{\nu j}$=220 $\gev$.
The PDF uncertainties were also estimated using 
the ZEUS NLO fit \cite{zeus_nlo_fit} with similar results.
\end{itemize}
Other uncertainties include jet-position reconstruction
and luminosity determination, which were small compared to those given above.
The overall systematic uncertainties 
on the background expectations were obtained by
summing the contributions from all these sources in quadrature,
and are shown in Figs.~\ref{cc_mass_posi}b and \ref{cc_mass_elec}b
as the shaded bands.


\subsubsection{Significance analysis}

The same significance analysis as applied to the $M_{ej}$ spectra
was performed on the $M_{\nu j}$ spectra. The results are
summarized in Table \ref{tab:many_experiment_cc}.
The observations are again
compatible with the SM expectation, and there is no evidence for
a narrow resonance in the $\nu q$ channel.

\section{Limits on the production of resonant states}

Since no evidence was found for a narrow resonance in the
lepton-jet mass spectra, limits were set on the Yukawa coupling
$\lambda$ and the production cross section
of the eight types of resonant states listed in Table
\ref{tab_resst} with masses below $\sqrt{s}$.
The resonance decay width was assumed to be small,
and the $s$-channel production was assumed to be
dominant, so that the resonance exchange and 
interference contributions were neglected.

The limits were set using a likelihood technique
involving the observables
 $M_{ljs}$ and  $\cos\theta^*$.  
The variable $M_{ljs}$ is
the invariant mass of the lepton-jets system and was calculated from:
\begin{equation}
\notag
M_{ljs}   =  \sqrt{2E_e{(E+P_Z)}_{ljs}} ~~,
\end{equation}
where $E_e=27.5\gev$ is the electron beam energy,
${(E+P_Z)}_{ljs}$ was determined
using the lepton and all jets in the event
satisfying $P_T^j>15\gev$ ($P_T^j>10\gev$ in $ep \to \nu X$
channel) and $\eta_j<3$
\cite{epj:c16:253,pr:d63:052002}. 

The $M_{ljs}$ method rather than the $M_{lj}$ method
was used in the limit setting because it has better resolution
on the
reconstructed resonance mass for $LQ \to eq$ events.
Using the PYTHIA MC described in Section
\ref{mc_simulation}, it was found that 
the resolution on $M_{ljs}$ 
for the $LQ \to eq$ events varied
from 3\% to 2\% as the mass varied from 150 to 290 GeV, while
that on $M_{lj}$ varied from 6\% to 4\%.
For the $LQ \to \nu q$ events, $M_{ljs}$ and $M_{lj}$ have similar 
resolutions.
On the other hand, the $M_{ljs}$ method assumes that
the final state consists only of the LQ decay products
and the proton remnant.  It is also affected by the
QED initial-state radiation (ISR) of a photon
from the incoming electron.
Therefore, the $M_{lj}$ method is more general and
better suited for a resonance search.
This is discussed in more detail elsewhere
\cite{epj:c16:253,pr:d63:052002}. 

In the limit-setting procedure, the width was 
assumed to be small compared to the detector resolution, so that
the narrow-width approximation (NWA) is valid.
In the NWA, the total cross section of
the production of a single state
is described at Born level\cite{pl:b191:442} by
\begin{equation}
\sigma^{NWA}=(J+1)\frac{\pi}{4s}\lambda^{2}q(x_{0}, M_{eq}^2) ,
\label{eqn_nwidth}
\end{equation}
where $q(x_{0}, M_{eq}^2)$ is the initial-state quark (or antiquark) 
momentum density in the proton
evaluated at $x_{0}=M_{eq}^{2}/s$ and virtuality 
scale $M_{eq}^2$,
and $J$ is the spin of the state. 

The effect of QED ISR on the resonance production cross section
was evaluated. It varies from
$-$4\% at resonance mass 100 $\gev$ to
$-$25\% at mass 300 $\gev$ and depends weakly on the
resonance type.
This was taken into account when setting the limits on $\lambda$.
The next-to-leading-order (NLO) QCD corrections,
the so-called $K$-factors,
including the vertex loop corrections,
gluon radiation from the leptoquark or the quark
and other higher-order diagrams, have been evaluated
for the scalar resonances 
\cite{zfp:c74:611,zfp:c75:453}.
The $K$-factor varied from 1.17 (1.17) to 1.15 (1.35)
for an $eu$ ($ed$) resonance with mass varying from 100 to 300 $\gev$.
However,
no calculation is available for the vector resonances,
so for consistency,
no NLO QCD correction was applied to either scalar or vector
resonance.

For each of the running periods listed in Table \ref{tab:lumi},
the $M_{ljs}$-$\cos\theta^*$ plane for
150 $<M_{ljs}<$ 320 $\gev$
was divided into bins, labelled $i$, 
for each of the $eq$ and $\nu q$ data 
samples.
For resonant states with $\nu q$ decays,
both $eq \to eX$ and $eq \to \nu X$ samples were used,
while for resonant states decaying only to $eq$,
only $eq \to eX$ samples were used.
The upper limit on the coupling strength,
$\lambda_{\rm limit}$,
as a function of $M_{eq}$, was obtained by solving
\footnote{With the Bayesian prior assumption of a 
uniform $\lambda^2$ distribution.}
\begin{equation}
\int_0^{\lambda_{\rm limit}^2}d\lambda^2
L(M_{eq},\lambda) = 
0.95 \int_0^{\infty}d\lambda^2
L(M_{eq}, \lambda),
\notag
\end{equation}
where 
$L$ is the product of
Poisson probabilities of all $\cos \theta^*$-$M_{ljs}$ bins 
convoluted with Gaussian distributions
for the main systematic uncertainties. The $L$ was calculated as
\begin{equation}
 L = \int_{-\infty}^{\infty} \prod_j d\delta_j 
   \frac{1}{\sqrt{2\pi}} e^{(-\delta_j^2/2)}
   \prod_i  e^{(-\mu'_i)} \frac{{\mu'_i}^{N_i}}{N_i!},
\notag
\end{equation}
   where $j$ denotes the source of systematic uncertainty and
$\delta_j$ corresponds to the variation of the
$j^{\rm th}$ systematic parameter in units of the nominal values
quoted in Sections \ref{sec:nc_systematic} and \ref{sec:cc_systematic}.
The index $i$ labels the bin in $\cos\theta^*$-$M_{ljs}$ and
$N_i$ gives the number of events observed in that bin.
The variable $\mu'_i$, which
denotes the expected number of events in bin $i$ after the effect
of the systematic variations, was calculated as
\[ \mu'_i = \mu_i\prod_j (1+\sigma_{ij})^{\delta_j}, \]
where $\mu_i$, which depends on the $M_{eq}$ and $\lambda$, is the number of
expected events in bin $i$ with no systematic variation
and $\sigma_{ij}$ gives the fractional variation of $\mu_i$ under the
nominal shift in the $j^{\rm th}$ systematic parameter.
This definition of $\mu'_i$ reduces to a linear dependence of
$\mu'_i$ on each $\delta_j$ when $\delta_j$ is small while avoiding
the possibility of $\mu'_i$ becoming negative which would arise if
$\mu'_i$ was defined as a linear function of the $\delta_j$'s.

\begin{figure}[tbph]
\begin{center}
\epsfig{figure=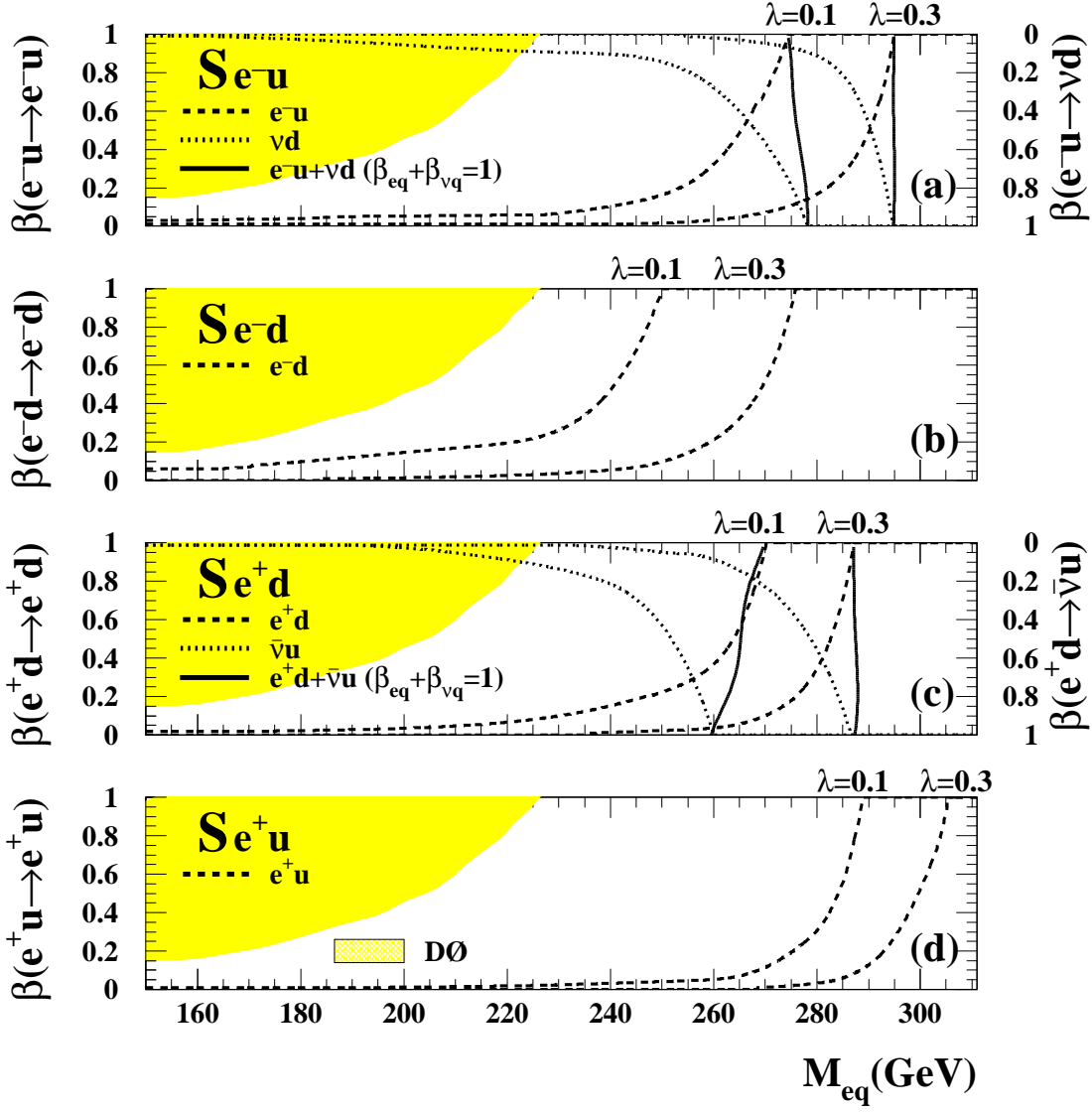,height=16.cm}
\end{center}
\caption{The constant-$\lambda$ limits as a function
of the branching ratios into $eq$ and $\nu q$ 
(shown on the left and right axes, respectively),
and of the resonance mass ($x$ axes), 
for the scalar resonant 
states listed in Table~\ref{tab_resst}.
The dotted line corresponds to limits set with only the 
$ep\to \nu X$ data set,
the dashed line with only the $ep\to eX$ data set;
the solid black line is the limit set using 
both data sets, assuming $\beta_{eq}+\beta_{\nu q}=1$.
For each limit curve, the area to the left of the curve 
is the excluded region.
Results for the resonant states 
are shown for constant limit of $\lambda=0.1$
and $\lambda=0.3$.
The shaded area in each plot 
shows the area excluded by the $\DO$ experiment.}
\label{br_scalar}
\end{figure}
\begin{figure}[tbph]
\begin{center}
\epsfig{figure=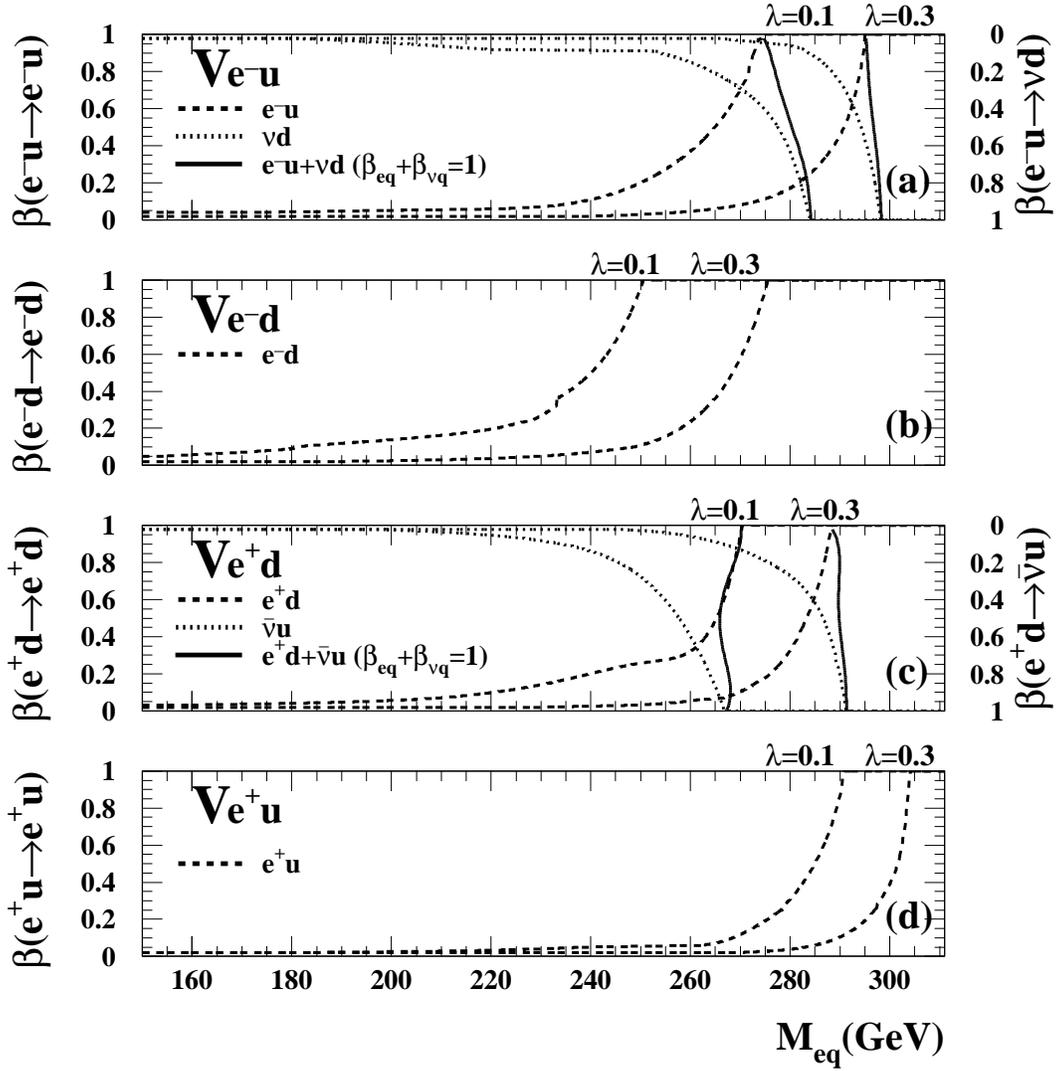,height=16.cm}
\end{center}
\caption{The constant-$\lambda$ limit contour for the vector resonant 
states listed in Table~\ref{tab_resst}.  For details, see the caption to
Fig. \ref{br_scalar}.}
\label{br_vector}
\end{figure}

Figure~\ref{br_scalar} shows the $\lambda$
limits as a function
of $\beta_{e q}$ and $\beta_{\nu q}$, 
the branching ratios for $LQ\to eq$ and $LQ\to \nu q$, respectively,
and as a function of the resonance mass.
These limits were obtained for the 
four scalar resonant states listed in Table~\ref{tab_resst}
and for $\lambda=0.1$ and $\lambda=0.3$, where
the latter ($\lambda \approx\sqrt{4\pi\alpha_{EW}}$)
corresponds to the electroweak (EW) coupling. 
The equivalent plots for vector resonant states are shown in 
Fig.~\ref{br_vector}.
For the $e^+u$ and $e^-d$  resonances 
$\nu q$ decays are forbidden
by charge conservation.
The $e^-u$ and $e^+d$ resonances 
can decay to either
$eq$ or $\nu q$. 
The $e q \! + \! \nu q$ limits, which assume 
$\beta_{\nu q}+\beta_{e q}=1$, 
are largely independent of the branching ratio,
and are typically in excess of 285 $\gev$ for the EW coupling.
The limits obtained using only the  
$eq$ (or the $\nu q$) data set
allow for decay modes other than $eq$ and $\nu q$,
so the $e q$ and $\nu q$ limits are applicable 
to a wider range of physics models than the combined 
$eq \! + \! \nu q$ results.
Similar limits have been presented by the H1 collaboration
for scalar $e^-d$ and $e^-u$ resonances~\cite{pl:b523:234}.

For comparison, the limits on scalar
resonances reported by the $\DO$ experiment~\cited0lq 
at the Tevatron are shown in Fig.~\ref{br_scalar}.
For a scalar resonant state with $\beta_{eq}$=1,
the $\DO$ and CDF~\cite{CDF_1stlq_results} limits are
$225\gev$ and $213\gev$, respectively, leading to a combined 
Tevatron limit of
$242\gev$~\cite{tevatron-combined-limit}.  These limits are
independent of both coupling and quark flavor, but
degrade (in the case of $\DO$ for example) 
to 204 GeV (98 $\gev$) as $\beta_{eq}$ decreases
from 100\% to 50\% (0\%).

The limits presented in Fig.~\ref{br_scalar}
apply to squarks with
$\rpviolate$ couplings to $eq$.
For example, the $\lambda$-limit contours on $S_{e^+d}$ with 
the decay $S_{e^+d} \to e^+d$
(the dashed curves in Fig.~\ref{br_scalar}c)
apply to $\tilde{u}_j$ squarks with coupling $\lambda_{1j1}$
and subsequent decay $\tilde{u}_j \to e^+d$, where the
subscript $j$ indicates the squark generation.
The limit curves on $S_{e^-u}$ with
decay $S_{e^-u} \to e^-u$
(the dashed curves in Fig.~\ref{br_scalar}a)
and with $S_{e^-u} \to \nu d$
(the dotted curves in the same plot)
apply to $\tilde{d}_j$ squarks with
coupling $\lambda_{11j}$ and subsequent
decays $\tilde{d}_j \to e^-u$ and $\tilde{d}_j \to \nu d$,
respectively.  With the $e^-u$ and $\nu d$ channels combined
and with $\beta_{e^-u}$ = $\beta_{\nu d}$ = 0.5$\beta$,
the limit for squark $\tilde{d}_j$ on $\lambda_{11j}\sqrt{\beta}$
is 0.1 for mass 276 $\gev$ and 0.3 for mass 295 $\gev$.
In this case, the $R_p$-conserving decay branching ratio of 
$\tilde{d}_j \to \chi_k q_j$ is 1-$\beta$,
where $\chi_k$ is the gaugino.

\begin{figure}[tbph]
\begin{center}
\epsfig{figure=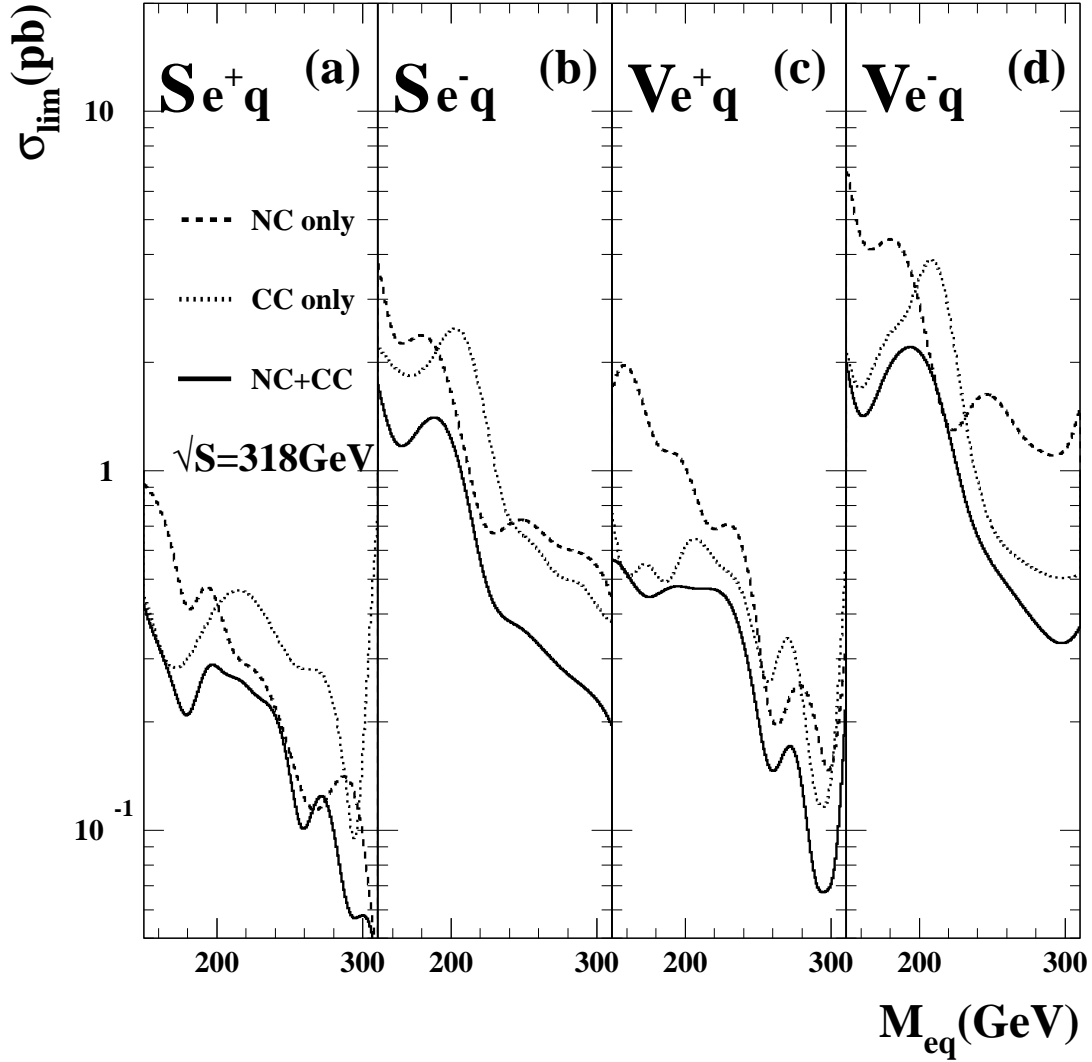,height=16.cm}
\end{center}
\caption{
(a,b) Limits on the total production cross section
for a narrow scalar resonant state and
(c,d) the corresponding limits for a narrow vector resonant state
as shown in Table \ref{tab_resst}.
Limits derived from NC (CC) data samples assume a branching ratio
$\beta_{eq}$ ($\beta_{\nu q}$) of 1/2, while the
combined $eq \!~ +~ \! \nu q$ limits assume branching
ratios of $\beta_{eq}=\beta_{\nu q}=1/2$.}
\label{resonance_cs}
\end{figure}

The limit on the resonant-state production cross section,
$\sigma_{\rm limit}$, was calculated by using the NWA
as shown in Eq.~(\ref{eqn_nwidth}) with $\lambda=\lambda_{\rm limit}$.
Figure \ref{resonance_cs} shows the results for the scalar
and vector resonant states.
Limits on the calculated production cross section
with only $eq$ data samples
(assuming $\beta_{eq}=$ 0.5), with only $\nu q$ data samples
(assuming $\beta_{\nu q}=$ 0.5) and with all the data samples combined
(assuming $\beta_{eq}=\beta_{\nu q}=$ 0.5) are shown 
as a function of $M_{eq}$.
Usually the limit becomes stricter after combining the
$eq$ and $\nu q$ data samples.

\section{Limits on BRW leptoquark model}
\label{sec_limits}

The two-dimensional ($M_{ljs}$-$\cos\theta^*$) 
likelihood method described above
was also used to set limits on the Yukawa coupling,
$\lambda$, of the BRW LQs.
The full LQ cross section was used, including
the $s$-channel and $u$-channel contributions 
and interference with DIS.
The difference of the limits thus obtained to
those obtained by using NWA is negligible 
below the LQ mass, $M_{LQ}$, of 280 $\gev$.

The coupling limits of
the 14 BRW LQs listed in Table~\ref{tab:brwlq}
were calculated as a function of $M_{LQ}$.
The presence of leptoquarks with $M_{LQ}>\sqrt{s}$ 
would affect the
observed mass spectra, particularly at high mass,
mainly due to $u$-channel exchange
and the interference effects.
Therefore, it is possible to extend the
limits beyond $\sqrt{s}$.  
Coupling limits were calculated
for masses up to $400\gev$.

\begin{figure}[ptb]
\centerline{\epsfxsize=11cm  \epsfysize=19cm
\epsfbox{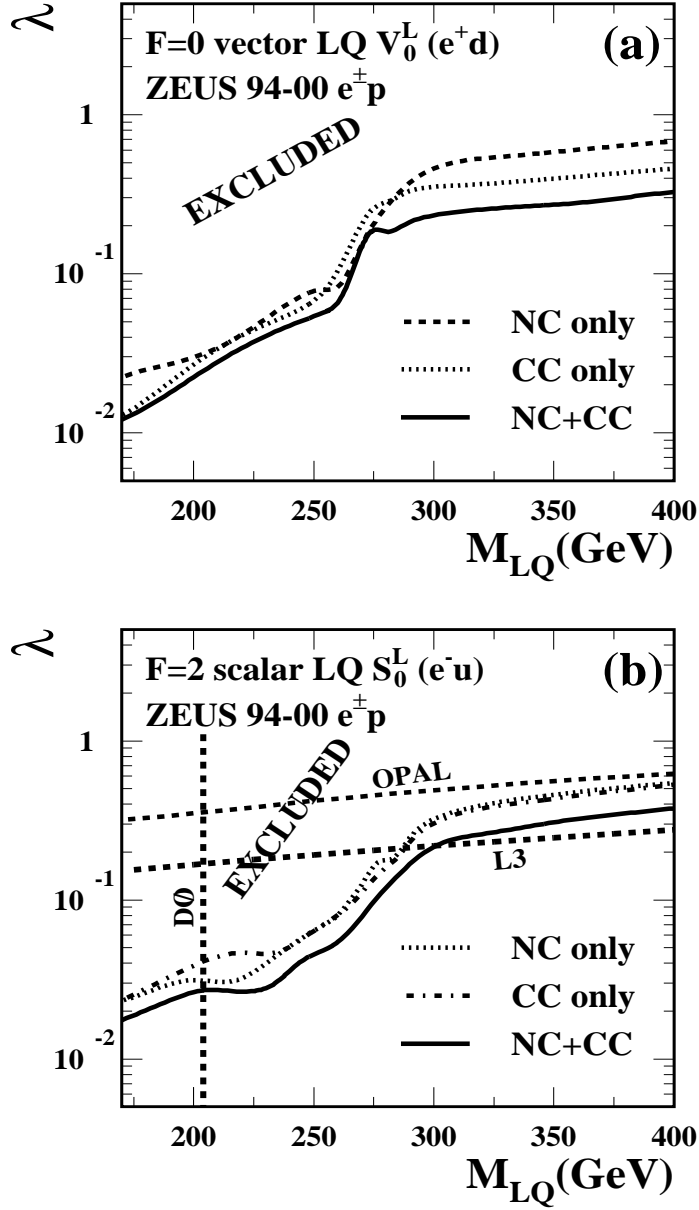}}
\vspace{-1.0cm}
\caption{Coupling limits as a function of LQ mass for (a) F=0 BRW LQ
$V^L_0$ and for (b) F=2 BRW LQ $S^L_0$.}
\label{llimitf2s0lf0v0l}
\end{figure}

The combined search in the $LQ\to eq$ and $LQ\to \nu q$ topologies,
when applicable,
produces more stringent
limits on leptoquark production.
An example is shown in Fig.~\ref{llimitf2s0lf0v0l}a
for the $V_0^L$ LQ produced through $e^+d$ fusion, 
which decays to $e^+d$ and $\bar{\nu} u$ with equal probability, 
and for the $S_0^L$ LQ produced through $e^-u$ fusion, 
which decays to $e^-u$ and $\nu d$ with equal probability.
The limits obtained with only the $eq$ data samples,
only the $\nu q$ data samples and the combined samples are shown
for each LQ. If a coupling strength 
$\lambda=0.3$ is assumed,
the production of an $V_0^L$ LQ is excluded up to a
mass of 284 (286)$\gev$ using $eq$ ($\nu q$) samples only,
while it is excluded up to 386$\gev$ with the combined samples.
For the $S_0^L$, the mass limit is 298 (301)$\gev$ using
$eq$ ($\nu q$) samples only and the combined mass limit is $351\gev$.

Also shown in Fig.~\ref{llimitf2s0lf0v0l}b
are the limits on the $S^L_0$ LQ from
$\DO$, OPAL~\cite{OPAL_1stlq_results} and L3~\cite{L3_1stlq_results}.
The most stringent limits in the high-mass region 
(above 200 GeV) from
the LEP experiments come from an indirect search through
the $t$-channel LQ exchange between the incoming electron
and positron.  

\begin{figure}[ptb]
\centerline{\epsfxsize=16cm  \epsfysize=16cm
\epsfbox{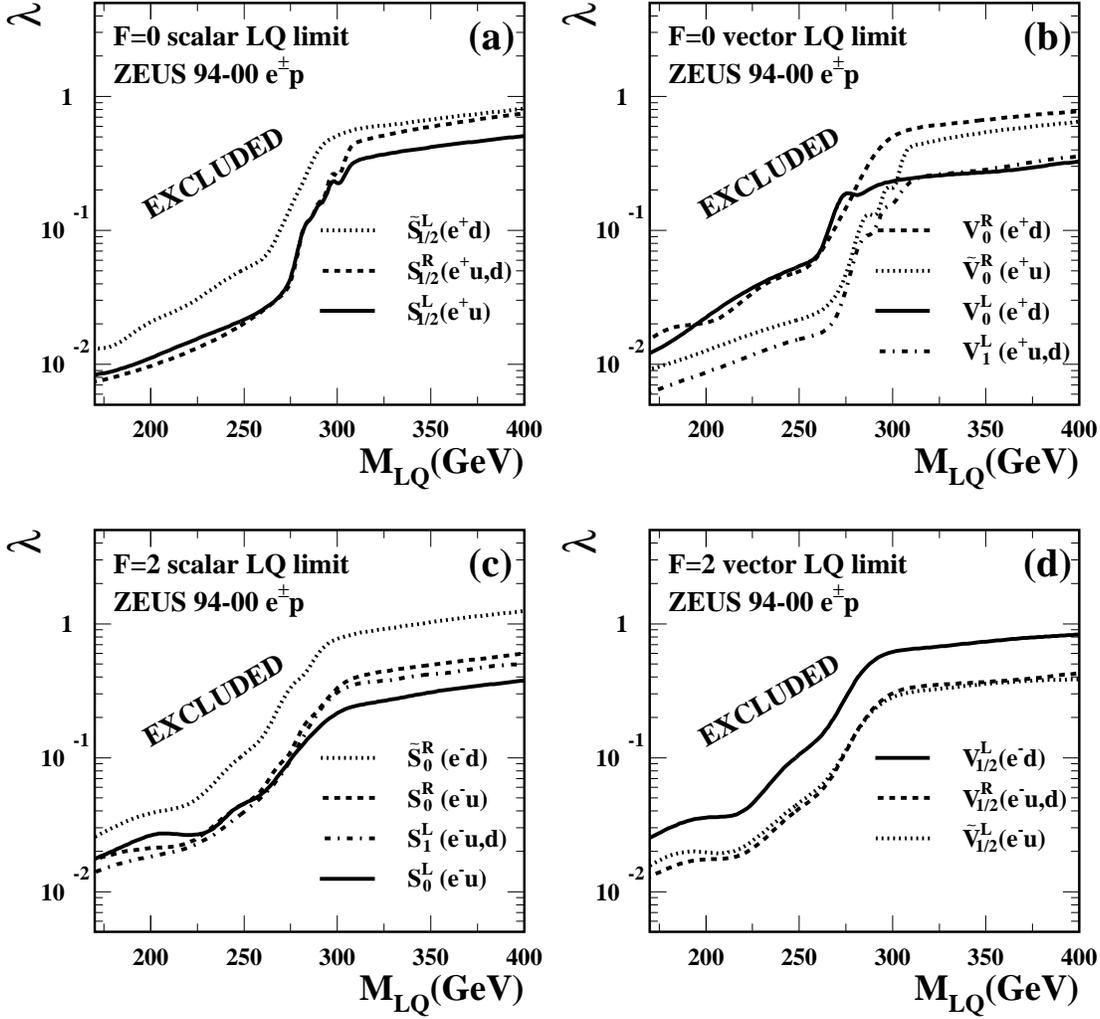}}
\caption{
Coupling limits as a function of LQ mass for 
scalar (a) and vector (b) F=0 BRW LQs and
for scalar (c) and vector (d) F=2 BRW LQs.
The areas above the curves are excluded.}
\label{llimitf0f2}
\end{figure}

Figure \ref{llimitf0f2} shows the coupling
limits on the scalar and vector BRW LQs with $F=0$ and $F=2$,
respectively, where $F=3B+L$ is the fermion number of the LQ
and $B$ and $L$ are the baryon and lepton numbers, respectively. 
The limits on $\lambda_{LQ}$ at $M_{LQ}=400\gev$
range from 0.3 to 1.0. 
In general, present results are significantly better than LEP
limits below 300 $\gev$ and comparable above 300 $\gev$.
The H1 collaboration has presented similar
limits on $F=0$~\cite{epj:c11:447} and
$F=2$~\cite{pl:b523:234} LQs.
The excluded mass regions
for BRW LQs with $\lambda=$~0.1 
and with $\lambda=$~0.3 are summarized in Table \ref{limit_table}.
They range from 248 to 290~$\gev$ for
$\lambda=$~0.1 and from 273 to 386~$\gev$
for $\lambda=$~0.3.

\section{Conclusions}

The total $ep$ data recorded by the ZEUS experiment at HERA
were used to search for the presence of a narrow-width
resonance decaying into 
a lepton and a jet,
with the final-state lepton being either an electron
or a neutrino. The data samples include
$16.7\pbi$ of $e^-p$ and
$114.8\pbi$ of $e^+p$ collisions.
No evidence was found for resonance
production, either in the $eq$ or $\nu q$ topology.
Limits were set on the coupling strength of 
resonant states that could decay in these topologies using a 
two-dimensional likelihood analysis. 
With the combined $eq$ and $\nu q$ topologies,
a scalar $e^-u$ ($e^+d$) resonant state is excluded up to
a mass of 275 (265) $\gev$ for coupling strength $\lambda=0.1$. 
The combined limits depend very weakly
on the resonance-decay branching ratio. Limits
on the coupling strength of 
Buchm$\ddot{\rm{u}}$ller-R$\ddot{\rm{u}}$ckl-Wyler-type 
leptoquarks with masses up to 400 $\gev$
are also presented.
The excluded mass regions depend on BRW LQ type
and the coupling. At $\lambda=$~0.1, they
range from 248 to 290~$\gev$, which are the most stringent
limits available to date.

\section*{Acknowledgements}
We thank the DESY Directorate for their strong support 
and encouragement,
and the HERA machine group for their diligent efforts.
We are grateful for the  support of the DESY computing
and network services. The design, construction and installation
of the ZEUS detector have been made possible by the ingenuity and effort
of many people who are not listed
as authors.


\begin{mcbibliography}{10}

\bibitem{pl:b191:442}
W.~Buchm\"uller, R.~R\"uckl and D.~Wyler,
\newblock Phys.\ Lett.{} {\bf B~191},~442~(1987).
\newblock Erratum in Phys.~Lett.~{\bf B~448}, 320 (1999)\relax
\relax
\bibitem{np:b397:3}
J.~Butterworth and H.~Dreiner,
\newblock Nucl.\ Phys.{} {\bf B~397},~3~(1993)\relax
\relax
\bibitem{epj:c16:253}
ZEUS \coll, J.~Breitweg \etal,
\newblock Eur.\ Phys.\ J.{} {\bf C~16},~253~(2000)\relax
\relax
\bibitem{pr:d63:052002}
ZEUS \coll, J.~Breitweg \etal,
\newblock Phys.\ Rev.{} {\bf D~63},~052002~(2001)\relax
\relax
\bibitem{zeus:1993:bluebook}
ZEUS \coll, U.~Holm~(ed.),
\newblock {\em The {ZEUS} Detector}.
\newblock Status Report (unpublished), DESY (1993),
\newblock available on
  \texttt{http://www-zeus.desy.de/bluebook/bluebook.html}\relax
\relax
\bibitem{nim:a279:290}
N.~Harnew \etal,
\newblock Nucl.\ Inst.\ Meth.{} {\bf A~279},~290~(1989)\relax
\relax
\bibitem{npps:b32:181}
B.~Foster \etal,
\newblock Nucl.\ Phys.\ Proc.\ Suppl.{} {\bf B~32},~181~(1993)\relax
\relax
\bibitem{nim:a338:254}
B.~Foster \etal,
\newblock Nucl.\ Inst.\ Meth.{} {\bf A~338},~254~(1994)\relax
\relax
\bibitem{nim:a309:77}
M.~Derrick \etal,
\newblock Nucl.\ Inst.\ Meth.{} {\bf A~309},~77~(1991)\relax
\relax
\bibitem{nim:a309:101}
A.~Andresen \etal,
\newblock Nucl.\ Inst.\ Meth.{} {\bf A~309},~101~(1991)\relax
\relax
\bibitem{nim:a321:356}
A.~Caldwell \etal,
\newblock Nucl.\ Inst.\ Meth.{} {\bf A~321},~356~(1992)\relax
\relax
\bibitem{nim:a336:23}
A.~Bernstein \etal,
\newblock Nucl.\ Inst.\ Meth.{} {\bf A~336},~23~(1993)\relax
\relax
\bibitem{desy-92-066}
J.~Andruszk\'ow \etal,
\newblock Preprint \mbox{DESY-92-066}, DESY, 1992\relax
\relax
\bibitem{zfp:c63:391}
ZEUS \coll, M.~Derrick \etal,
\newblock Z.\ Phys.{} {\bf C~63},~391~(1994)\relax
\relax
\bibitem{acpp:b32:2025}
J.~Andruszk\'ow \etal,
\newblock Acta Phys.\ Pol.{} {\bf B~32},~2025~(2001)\relax
\relax
\bibitem{cpc:135:238}
T.~Sj\"{o}strand,
\newblock Comp.\ Phys.\ Comm.{} {\bf 135},~238~(2001)\relax
\relax
\bibitem{pythia_lq61_hera}
C.~Friberg, E.~Norrbin and T.~Sj\"ostrand,
\newblock Phys. Lett.{} {\bf B~403},~329~(1997)\relax
\relax
\bibitem{cpc:69:155}
A.~Kwiatkowski, H.~Spiesberger and H.-J.~M\"ohring,
\newblock Comp.\ Phys.\ Comm.{} {\bf 69},~155~(1992)\relax
\relax
\bibitem{cpc:81:381}
K.~Charchula, G.A.~Schuler and H.~Spiesberger,
\newblock Comp.\ Phys.\ Comm.{} {\bf 81},~381~(1994)\relax
\relax
\bibitem{spi:www:djangoh11}
H.~Spiesberger,
\newblock {\em {\sc heracles} and {\sc djangoh}: Event Generation for $ep$
  Interactions at {HERA} Including Radiative Processes}, 1998,
\newblock available on \texttt{http://www.desy.de/\til
  hspiesb/djangoh.html}\relax
\relax
\bibitem{cpc:101:108}
G.~Ingelman, A.~Edin and J.~Rathsman,
\newblock Comp.\ Phys.\ Comm.{} {\bf 101},~108~(1997)\relax
\relax
\bibitem{pr:d55:1280}
H.L.~Lai \etal,
\newblock Phys.\ Rev.{} {\bf D~55},~1280~(1997)\relax
\relax
\bibitem{epj:c14:285}
M.~Botje,
\newblock Eur.\ Phys.\ J.{} {\bf C~14},~285~(2000)\relax
\relax
\bibitem{tech:cern-dd-ee-84-1}
R.~Brun et al.,
\newblock {\em {\sc geant3}},
\newblock Technical Report CERN-DD/EE/84-1, CERN, 1987\relax
\relax
\bibitem{zfp:c74:207}
ZEUS \coll, J.~Breitweg \etal,
\newblock Z.\ Phys.{} {\bf C~74},~207~(1997)\relax
\relax
\bibitem{np:b406:187}
S.~Catani \etal,
\newblock Nucl.\ Phys.{} {\bf B~406},~187~(1993)\relax
\relax
\bibitem{pr:d48:3160}
S.D.~Ellis and D.E.~Soper,
\newblock Phys.\ Rev.{} {\bf D~48},~3160~(1993)\relax
\relax
\bibitem{epj:c12:411}
ZEUS \coll, J.~Breitweg \etal,
\newblock Eur.\ Phys.\ J.{} {\bf C~12},~411~(2000)\relax
\relax
\bibitem{pl:b539:197}
ZEUS \coll, S.~Chekanov \etal,
\newblock Phys.\ Lett.{} {\bf B~539},~197~(2002)\relax
\relax
\bibitem{zeus_nlo_fit}
ZEUS \coll, S.~Chekanov \etal,
\newblock Phys.\ Rev.{} {\bf D~67},~012007~(2003)\relax
\relax
\bibitem{zfp:c74:611}
T.~Plehn \etal,
\newblock Z.\ Phys.{} {\bf C~74},~611~(1997)\relax
\relax
\bibitem{zfp:c75:453}
Z.~Kunszt and W.J.~Stirling,
\newblock Z.\ Phys.{} {\bf C~75},~453~(1997)\relax
\relax
\bibitem{pl:b523:234}
H1 \coll, C.~Adloff \etal,
\newblock Phys.\ Lett.{} {\bf B~523},~234~(2001)\relax
\relax
\bibitem{DO_1stlq_eejets}
D{\O} \coll, B.~Abbott \etal,
\newblock Phys.\ Rev.\ Lett.{} {\bf 79},~4321~(1997)\relax
\relax
\bibitem{DO_1stlq_results}
D{\O} \coll, B.~Abbott \etal,
\newblock Phys.\ Rev.\ Lett.{} {\bf 80},~2051~(1998)\relax
\relax
\bibitem{DO_1stlq_sum}
D{\O} \coll, V.M.~Abazov \etal,
\newblock Phys.\ Rev.{} {\bf D~64},~092004~(2001)\relax
\relax
\bibitem{DO_1stlq_nunujets}
D{\O} \coll, V.M.~Abazov \etal,
\newblock Phys.\ Rev.\ Lett.{} {\bf 88},~191801~(2002)\relax
\relax
\bibitem{CDF_1stlq_results}
CDF \coll, F.~Abe \etal,
\newblock Phys.\ Rev.\ Lett.{} {\bf 79},~4327~(1997)\relax
\relax
\bibitem{tevatron-combined-limit}
Leptoquark Limits Combination Working Group, C.~Grosso-Pilcher, G.~Landsberg
  and M.~Paterno,
\newblock Preprint \mbox{hep-ex/9810015}\relax
\relax
\bibitem{OPAL_1stlq_results}
OPAL \coll, G.~Abbiendi \etal,
\newblock Eur. Phys. J.{} {\bf C~6},~1~(1999)\relax
\relax
\bibitem{L3_1stlq_results}
L3 \coll, M. Acciari \etal,
\newblock Phys.\ Lett.{} {\bf B~489},~81~(2000)\relax
\relax
\bibitem{epj:c11:447}
H1 \coll, C.~Adloff \etal,
\newblock Eur.\ Phys.\ J.{} {\bf C~11},~447~(1999).
\newblock Erratum in Eur.~Phys.~J.~{\bf C~14}, 553 (2000)\relax
\relax
\bibitem{zfp:c46:679}
A.~Djouadi \etal,
\newblock Z.\ Phys.{} {\bf C~46},~679~(1990)\relax
\relax
\end{mcbibliography}
\clearpage
\providecommand{\etal}{et al.\xspace}
\providecommand{\coll}{Coll.\xspace}
\catcode`\@=11
\def\@bibitem#1{%
\ifmc@bstsupport
  \mc@iftail{#1}%
    {;\newline\ignorespaces}%
    {\ifmc@first\else.\fi\orig@bibitem{#1}}
  \mc@firstfalse
\else
  \mc@iftail{#1}%
    {\ignorespaces}%
    {\orig@bibitem{#1}}%
\fi}%
\catcode`\@=12

\clearpage
\newpage
\begin{table}[ptb]
\begin{center}
\begin{tabular}
[c]{|c|c|c|c|c|c|c|}
\hline
 Period  & $E_p$ (GeV)  & $E_e$ (GeV) & $\sqrt{s}$ (GeV) & $e$ charge 
& luminosity ($\pbi$)  & $\delta_{lumi}$\\ 
 \hline \hline
94-97    & 820  & 27.5  & 300 & $e^+$  & 48.5 
   & $\pm$1.6\% \\ \hline
98-99    & 920  & 27.5  & 318 & $e^-$  & 16.7
   & $\pm$1.8\% \\ \hline
99-00    & 920  & 27.5  & 318 & $e^+$  & 66.3
   & $\pm$2.25\% \\ \hline
\end{tabular}
\end{center}
\caption{The characteristics of 
the three data samples used for the present study;
$\delta_{lumi}$ is the uncertainty of the measured luminosity.
}
\label{tab:lumi}
\end{table}

\begin{table}[ptb]
\begin{center}
\begin{tabular}[c]{|c||c|c||c|c|}\hline
     & \multicolumn{2}{|c||}{No $\cos\theta^*$ cut}
     & \multicolumn{2}{|c|}{$\cos\theta^*<$0.4} \\ \hline
     &  $e^+p$ &  $e^-p$ & $e^+p$ & $e^-p$ \\ \hline\hline
$P_{min}$ & 0.26\%   & 0.032\%  & 1.9\% & 4.3\% \\ \hline
$M_{ej}$($\gev$)  & 121 & 151   & 158   & 151   \\ \hline
$N_{obs}$ & 2575     & 305      & 73    & 27    \\ \hline
$\mu_{SM}$& 2436     & 249      & 56.4  & 18.8  \\ \hline
$F$       & 35\%     & 6\%      & 41\%  & 56\%  \\ \hline
\end{tabular}
\caption{Results of the significance analysis on
the $M_{ej}$ spectra:
$P_{min}$ is the minimum probability along the 
$M_{ej}$ spectra, as defined in Eq.~(\ref{poisson_more});
$N_{obs}$ is the number of observed events
at the corresponding $M_{ej}$;
$\mu_{SM}$ is the expectation from the SM background;
$F$ is the fraction of the simulated experiments 
with the minimum probability $P<P_{min}$.}
\label{tab:many_experiment_nc}
\end{center}
\end{table}

\begin{table}[ptb]
\begin{center}
\begin{tabular}[c]{|c||c|c||c|c|}\hline
     & \multicolumn{2}{|c||}{No $\cos\theta^*$ cut}
     & \multicolumn{2}{|c|}{$\cos\theta^*<$0.4} \\ \hline
     &  $e^+p$ &  $e^-p$ & $e^+p$ & $e^-p$ \\ \hline\hline
$P_{min}$ & 1.8\%    &   9.9\%  & 1.1\% & 24.5\% \\ \hline
$M_{\nu j}$($\gev$) & 246  & 225   &  269  & 217    \\ \hline
$N_{obs}$ & 10       & 22       & 2     & 9      \\ \hline
$\mu_{SM}$& 4.5      & 16.2     & 0.16  & 6.8  \\ \hline
$F$   & 76\%     & 77\%     & 24\%  & 94\%  \\ \hline
\end{tabular}
\caption{Results of the significance analysis on
the $M_{\nu j}$ spectra.  For details, see the caption to
Table \ref{tab:many_experiment_nc}.}
\label{tab:many_experiment_cc}
\end{center}
\end{table}

\begin{table}[ptb]
\begin{center}
\begin{tabular}[c]{|c|c|c||c|c|c|}\hline
\multicolumn{3}{|c|}{Scalar} & \multicolumn{3}{|c|}{Vector} \\ \hline
Resonance   & Charge   & Decay   & Resonance & Charge & Decay \\ \hline
$S_{e^+ u}$ & 5/3 & $e^+ u$ & $V_{e^+ u}$ & 5/3 & $e^+ u$ \\
$S_{e^+ d}$ & 2/3 & $e^+ d$ & $V_{e^+ d}$ & 2/3 & $e^+ d$ \\ 
 & & $\bar \nu u$   & & & $\bar \nu u$ \\ \hline \hline
$S_{e^- u}$ & -1/3 & $e^- u$ & $V_{e^- u}$ & -1/3 & $e^- u$ \\
& & $\nu d$ &  &  &$\nu  d$ \\
$S_{e^-  d}$ & -4/3 & $e^-  d$ & $V_{e^- d}$ & -4/3 & $e^- d$ \\ \hline
\end{tabular}
\end{center}
\caption{First-generation scalar and vector
resonant states that can be produced in $e^{\pm}p$ scattering.
The top half of the table lists color-triplet 
states with fermion number
$F=L+3B=0$, while the bottom half lists those with $F=2$. The left and
right sets of columns list scalars and vectors, respectively.
}
\label{tab_resst}
\end{table}

\begin{table}[hbpt]
\begin{displaymath}
\begin{tabular}{c|crlcc}

LQ species & $q$ & Production & Decay & Branching ratio & Coupling
 \\ \hline
$S_{1/2}^L$ & 5/3 & $e^+_R u_R$ & $e^+ u$ & 1 & $\lambda_L$ \\ 
$S_{1/2}^R$ & 5/3 & $e^+_L u_L $ & $e^+u$ & 1 & $\lambda_R$ \\ 
& 2/3 & $e^+_L d_L $ & $e^+ d$ & 1 & $-\lambda_R$ \\ 
${\tilde S}_{1/2}^L$ & -2/3 & $e^+_R d_R $ & $e^+d$ & 1 & $\lambda_L$ \\ 
$V_0^L$ & 2/3 & $e^+_R d_L $ & $e^+ d$ & 1/2 & $\lambda_L$ \\ 
&  & 
$$
& $\bar{\nu}_e u$ & 1/2 & $\lambda_L$\\ 
$V_0^R$ & 2/3 & $e^+_L d_R $ & $e^+ d$ & 1 & $\lambda_R$ \\ 
${\tilde V}_0^R$ & 5/3 & $e^+_L u_R $ & $e^+u$ & 1 & $\lambda_R$ \\ 
$V_1^L$ & 5/3 & $e^+_Ru_L $ & $e^+ u$ & 1 & $\sqrt{2}\lambda_L$\\ 
& 2/3 & $e^+_R d_L $ & $e^+ d$ & 1/2 & $-\lambda_L$\\ 
&  & 
$$
& $\bar{\nu}_e u$ & 1/2 & $\lambda_L$ \\ \hline
$S^L_0$          &-1/3& $e^-_Lu_L$   & $e^-u$  &1/2& $\lambda_L$\\
                 &    & $        $   &$\nu_ed$ &1/2& $-\lambda_L$\\
$S^R_0$          &-1/3& $e^-_Ru_R$   & $e^-u$  &1  & $\lambda_R$\\
$\tilde{S}^R_0$  &-4/3& $e^-_Rd_R$   & $e^-d$     & 1 & $\lambda_R$\\
$S^L_1$          &-1/3& $e^-_Lu_L$   & $e^-u$     &1/2& $-\lambda_L$\\
                 &    &            &$\nu_ed$ &1/2& $-\lambda_L$\\
                 &-4/3& $e^-_Ld_L$   &$e^-d$& 1 &$-\sqrt{2}\lambda_L$ \\
$V^L_{1/2}$      &-4/3& $e^-_Ld_R$   &$e^-d$      & 1 & $\lambda_L$ \\
$V^R_{1/2}$      &-4/3& $e^-_Rd_L$   &$e^-d$     & 1 & $\lambda_R$ \\
                 &-1/3& $e^-_Ru_L$   &$e^-u$     & 1 & $\lambda_R$ \\
$\tilde{V}^L_{1/2}$&-1/3&$e^-_Lu_R$  &$e^-u$     & 1 & $\lambda_L$ 
\\ \hline
\end{tabular}
\end{displaymath}
\caption{The $F=0$ (upper part) 
and $F=2$ (lower part) 
leptoquark species
of the Buchm$\ddot{{u}}$ller-R$\ddot{{u}}$ckl-Wyler
model \protect\cite{pl:b191:442}
and the corresponding couplings. Those LQs that couple only to
neutrino and quark and therefore could not be produced at HERA
are not listed.
The LQ species are classified according to their spin ($S$
for scalar and $V$ for vector), their chirality ($L$ or $R$) and 
their weak
isospin ($0,1/2,1$). The leptoquarks ${\tilde S}$ and ${\tilde V}$ 
differ by
two units of hypercharge from $S$ and $V$, respectively. In addition,
 the
electric charge $q$ of the leptoquarks, the production channel,
 as well as
their allowed decay channels assuming lepton-flavor conservation
are displayed.
The nomenclature follows the Aachen convention
\protect\cite{zfp:c46:679}.}
\label{tab:brwlq}
\end{table}

\begin{table}[h]
\begin{center}{\small
\begin{tabular}{|c||c|c|c|c|c|c|c|}
\hline
LQ Type (F=0)  &  $V_0^L$       & $V_0^R$     
         &      $\tilde{V}^R_0$   & $V_1^L$ & 
           $S_{1/2}^L$  & $S_{1/2}^R$ & $\tilde{S}_{1/2}^L$  \\
\hline\hline
M(GeV)($\lambda$=0.1)  &   266 &   268&   282 &   290&   282 &   282&   269 \\ \hline
M(GeV)($\lambda$=0.3)  &   386 &   287&   305 &   367&   308 &   303&   286 \\ \hline\hline
LQ Type (F=2)  &  $S_0^L$    & $S_0^R$     
         & $\tilde{S}^R_0$   &  $S_1^L$ & 
           $V_{1/2}^L$  & $V_{1/2}^R$ & $\tilde{V}_{1/2}^L$  \\
\hline\hline
M(GeV)($\lambda$=0.1) &   276  &   273&   248 &   275&   248 &   274&   273 \\ \hline
M(GeV)($\lambda$=0.3) &   351  &   298&   273 &   300&   277 &   302&   313 \\ \hline\hline
\end{tabular}\caption{Mass limit of the 14 BRW LQs at 
$\lambda$=0.1 and 0.3.}
\label{limit_table}
}\end{center}
\end{table}


\clearpage
\newpage


\clearpage

\clearpage


%
%
\end{document}